%% file: neurips_2022.tex
\documentclass{article}

\pdfoutput=1

\PassOptionsToPackage{numbers, compress}{natbib}

\usepackage[final]{neurips_2022}

\usepackage[utf8]{inputenc} 
\usepackage[T1]{fontenc}    
\usepackage[hidelinks]{hyperref}       
\usepackage{enumitem}
\usepackage{url}            
\usepackage{xspace}
\usepackage{booktabs}       
\usepackage{amsfonts}       
\usepackage{nicefrac}       
\usepackage{microtype}      
\usepackage{xcolor}         
\usepackage[noend]{algpseudocode}
\usepackage{algorithm}
\usepackage{multirow}
\usepackage{multicol}
\usepackage{amsmath}
\usepackage{amssymb}
\usepackage{amsthm}
\usepackage{graphicx}
\usepackage{subcaption}
\usepackage{wrapfig}
\usepackage{mathtools}
\usepackage[capitalise, noabbrev]{cleveref}

\title{Infinite Recommendation Networks: A Data-Centric Approach}

\author{%
  Noveen Sachdeva$^\dagger$ \\
  \And Mehak Preet Dhaliwal$^\dagger$  \\
    \And Carole-Jean Wu$^\ddagger$ \\
      \And Julian McAuley$^\dagger$ \\
        \AND
        {}\vspace{-0.7cm}\\      
    University of California, San Diego$^\dagger$\quad Meta AI$^\ddagger$\\[1ex]
    \texttt{\{nosachde,mdhaliwal,jmcauley\}@ucsd.edu}\\[.5ex] \texttt{carolejeanwu@meta.com}
}

\begin{document}

\include{general_definitions}
\include{recsys_definitions}

\maketitle

\begin{abstract}
    We leverage the Neural Tangent Kernel and its equivalence to training infinitely-wide neural networks to devise \model: an autoencoder with infinitely-wide bottleneck layers. The outcome is a highly expressive yet simplistic recommendation model with a single hyper-parameter and a closed-form solution. Leveraging \model's simplicity, we also develop \sampler for synthesizing tiny, high-fidelity data summaries which distill the most important knowledge from the extremely large and sparse user-item interaction matrix for efficient and accurate subsequent data-usage like model training, inference, architecture search, \etc This takes a data-centric approach to recommendation, where we aim to improve the quality of logged user-feedback data for subsequent modeling, independent of the learning algorithm. We particularly utilize the concept of differentiable Gumbel-sampling to handle the inherent data heterogeneity, sparsity, and semi-structuredness, while being scalable to datasets with hundreds of millions of user-item interactions. Both of our proposed approaches significantly outperform their respective state-of-the-art and when used together, we observe $96-105$\% of \model's performance on the full dataset with as little as $0.1$\% of the original dataset size, leading us to explore the counter-intuitive question: \emph{Is more data what you need for better recommendation?}
\end{abstract}

\section{Introduction}
The Neural Tangent Kernel (NTK) has recently advanced the theoretical understanding of how neural networks learn \cite{finite_vs_infinite_3, ntk}. Notably, performing Kernelized Ridge Regression (KRR) with the NTK has been shown to be equivalent to training infinitely-wide neural networks for an infinite number of SGD steps. Owing to KRR's closed-form solution,
these networks can be trained in a fast and efficient manner whilst not sacrificing expressivity.
While the application of infinite neural networks is being explored in various learning problems \cite{matrix_completion_ntk, ntk_small_data}, detailed comparative analyses demonstrate that deep, finite-width networks tend to perform significantly better than their infinite-width counterparts for a variety of standard computer-vision tasks \cite{finite_vs_infinite_2}. 

On the contrary, for recommendation tasks, there always has been a debate of linear \vs non-linear networks \cite{linear_or_non_linear, rethinking_ntk}, along with the importance of increasing the width \vs depth of the network \cite{wide_and_deep,naumov:dlrm}. At a high level, the general conclusion is that a well-tuned, wide and linear network can outperform shallow and deep neural networks for recommendation \cite{mf_vs_ncf}. We extend this debate by introducing our model \model, an autoencoder with infinitely wide bottleneck layers and examine its behavior on the recommendation task. Our evaluation demonstrates significantly improved results over state-of-the-art (SoTA) models across various datasets and evaluation metrics. 


A rising challenge in recommender systems research has been the increased cost of training models on massive datasets which can involve billions of user-item interaction logs, in terms of time, computational resources, as well as the downstream carbon footprint. Moreover, despite anonymization efforts, privacy risks have been associated with publicly released user data \cite{netflix_deanon}.
To this end, we further explore recommendation from a data-centric viewpoint \cite{ng_data_centric}, which we loosely define as: 

\begin{definition} \label{def:data_centric}
    {\normalfont \textbf{(Data-centric AI)}} Systematic methods for improving the collected data's quality, thereby shifting the focus from merely acquiring large quantities of data; implicitly helping in a learning algorithm's generalization, scalability, and eco-sustainability.
\end{definition}

Previous work on data-centric techniques generally involve constructing terse data summaries of large datasets, and has focused on domains with continuous, real-valued features such as images \cite{zhao_dsa, kip_conv}, which are arguably more amenable to data optimization approaches. Due to the heterogeneity, sparsity, and semi-structuredness of recommendation data, such methods are not directly applicable. Common approaches for scaling down such recommendation datasets typically include heuristics such as random, head-user, or k-core sampling. More principled approaches include coreset construction \cite{wsdm22} that focus on optimizing for \emph{picking} the set of data-points that are the most representative from a given dataset, and are generally shown to out-perform various heuristic sampling strategies. However, \emph{synthesizing} informative summaries for recommendation datasets still remains a challenge.

Consequently, we propose \sampler, a data distillation framework for collaborative filtering (CF) datasets that utilizes \model in a bilevel optimization objective to create highly compressed, informative, and generic data summaries. \sampler employs efficient multi-step differentiable Gumbel-sampling \cite{gumbel} at each step of the optimization to encompass the heterogeneity, sparsity, and semi-structuredness of recommendation data.
We further provide an analysis of the denoising effect observed when training the model on the synthesized versus the full dataset.

To summarize, \emph{in this paper}, we make the following contributions:
\begin{itemize}[leftmargin=.3in]
    \item We develop \model: an infinite-width autoencoder for recommendation, that aims to reconstruct the incomplete preferences in a user's item consumption sequence. We demonstrate its efficacy on four datasets, and demonstrate that \model outperforms complicated SoTA models with only a single fully-connected layer, closed-form optimization, and a single hyper-parameter. We believe our work to be the first to demonstrate that an infinite-width network can outperform their finite-width SoTA counterparts for practical scenarios like recommendation.
    
    \item We subsequently develop \sampler: a novel data distillation framework that can synthesize tiny yet accurate data summaries for a variety of modeling applications. We empirically demonstrate similar performance of models trained on the full dataset versus training the same models on $2-3$ orders smaller data summaries synthesized by \sampler.
    Notably, \sampler and \model are complementary for each other's practicality, as \model's closed-form solution enables \sampler to scale to datasets with hundreds of millions of interactions; whereas, \sampler's succinct data summaries help improving \model's restrictive training complexity, and achieving SoTA performance when trained on these tiny data summaries.
    
    \item Finally, we also note that \sampler has a strong data denoising effect, validated with a counter-intuitive observation --- if there is noise in the original data, models trained on \emph{less} data synthesized by \sampler can be better than the same model trained on the \emph{entire} original dataset. 
    Such observations, along with the strong data compression results attained by \sampler, reinforce our data-centric viewpoint to recommendation, encouraging the community to think more about the quality of collected data, rather than its quantity. 
\end{itemize}


\section{Related Work}
\paragraph{Autoencoders for recommendation.} Recent approaches to implicit feedback recommendation involve building models that can re-construct an incomplete user preference list using autoencoders \cite{mvae, ease, svae, macridvae}. The CDAE model \cite{cdae} first introduced this idea and used a standard denoising autoencoder for recommending new items to users. MVAE \cite{mvae} later extended this idea to use variational autoencoders, and provided a principled approach to perform variational inference for this model architecture. More recently, EASE \cite{ease} proposed using a shallow autoencoder and estimates only an item-item similarity matrix by performing ordinary least squares regression on the relevance matrix, resulting in closed-form optimization.

\paragraph{Infinite neural networks.} The Neural Tangent Kernel (NTK) \cite{ntk} has gained significant attention because of its equivalence to training infinitely-wide neural networks by performing KRR with the NTK. Recent work also demonstrated the double-descent risk curve \cite{double_descent} that extends the classical regime of train \vs test error for overparameterized neural networks, and plays a crucial role in the generalization capabilities of such infinite neural networks. However, despite this emerging promise of utilizing NTK for learning problems, detailed comparative analyses \cite{finite_vs_infinite_1, finite_vs_infinite_2, finite_vs_infinite_3} for computer vision tasks demonstrate that finite-width networks are still significantly more accurate than infinite-width ones. Looking at recommendation systems, \cite{rethinking_ntk} performed a theoretical comparison between Matrix Factorization (MF) and Neural MF \cite{neural_mf} by studying their expressivity in the infinite-width limit, comparing the NTK of both of these algorithms. Notably, their settings involved the typical (user-ID, item-ID) input to the recommendation model, and observed results that were equivalent to a random predictor. \cite{matrix_completion_ntk} performed a similar study that performed matrix completion using the NTK of a single layer fully-connected neural network, but assumed meaningful feature-priors available beforehand.

\paragraph{Data sampling \& distillation.} Data sampling is ubiquitous --- sampling negatives while contrastive learning \cite{negative_contrastive, negative_contrastive_2}, sampling large datasets for fast experimentation \cite{wsdm22}, sampling for evaluating expensive metrics \cite{sampled_metrics}, \etc In this paper, we primarily focus on sampling at the dataset level, principled approaches of which can be categorized into: (1) coreset construction methods that aim to \emph{pick} the most useful datapoints for subsequent model training \cite{bilevel_coresets, grad_match, selcon_coreset, coreset_scale}. These methods typically assume the availability of a submodular set-function $f : \mathbf{V} \mapsto \mathbb{R}_+ ~ \forall ~ \mathbf{V} \subseteq \mathbf{X}$ for a given dataset $\mathbf{X}$ (see \cite{bilmes_submodularity} for a systematic review on submodularity), and use this set-function $f$ as a proxy to guide the search for the most informative subset; and (2) dataset distillation: instead of picking the most informative data-points, dataset distillation techniques aim to \emph{synthesize} data-points that can accurately distill the knowledge from the entire dataset into a small, synthetic data summary. Originally proposed in \cite{dd_orig}, the authors built upon the notions of gradient-based hyper-parameter optimization \cite{maclaurin} to synthesize representative images for model training. Subsequently, a series of works \cite{zhao_dc, zhao_dsa, kip, kip_conv} propose various subtle modifications to the original framework, for improving the sample-complexities of models trained on data synthesized using their algorithms. Note that such distillation techniques focused on continuous data like images, which are trivial to optimize in the original data-distillation framework. More recently, \cite{graph_distill} proposed a distillation technique for synthesizing fake graphs, but also assumed to have innate node representations available beforehand, prohibiting their method's application for recommendation data.

\section{\model: Infinite-width Autoencoders for Recommendation} \label{sec:inf_ae}

\paragraph{Notation.} Given a recommendation dataset $\mathcal{D} \coloneqq \left\{ (\text{user}_i, \text{item}_i, \text{relevance}_i) \right\}_{i=1}^{n}$ consisting of $n$ user-item interactions defined over a set of users $\mathcal{U}$, set of items $\mathcal{I}$, and operating over a binary relevance score (implicit feedback); we aim to best model user preferences for item recommendation. The given dataset $\mathcal{D}$ can also be viewed as an interaction matrix, $X \in \mathbb{R}^{|\mathcal{U}| \times |\mathcal{I}|}$ where each entry $X_{u, i}$ either represents the observed relevance for item $i$ by user $u$ or is missing. Note that $X$ is typically extremely sparse, \ie, $n \ll |\mathcal{U}| \times |\mathcal{I}|$. More formally, we define the problem of recommendation as accurate likelihood modeling of $\text{P}(i ~|~ u, \mathcal{D}) ~ \forall u \in \mathcal{U}, ~ \forall i \in \mathcal{I}$.

\paragraph{Model.} \model takes an autoencoder approach to recommendation, where the all of the bottleneck layers are infinitely-wide. Firstly, to make the original bi-variate problem of which \emph{item} to recommend to which \emph{user} more amenable for autoencoders, we make a simplifying assumption that a user can be represented only by their historic interactions with items, \ie, the much larger set of users lie in the linear span of items. This gives us two kinds of modeling advantages: (1) we no longer have to find a unique latent representation of users; and (2) allows \model to be trivially applicable for any user not in $\mathcal{U}$. More specifically, for a given user $u$, we represent it by the sparse, bag-of-words vector of historical interactions with items $X_u \in \mathbb{R}^{|\mathcal{I}|}$, which is simply the $u^{\text{th}}$ row in $X$. We then employ the Neural Tangent Kernel (NTK) \cite{ntk} of an autoencoder that takes the bag-of-items representation of users as input and aims to reconstruct it. Due to the infinite-width correspondence \cite{ntk, finite_vs_infinite_3}, performing Kernelized Ridge Regression (KRR) with the estimated NTK is equivalent to training an infinitely-wide autoencoder for an infinite number of SGD steps.
More formally, given the NTK, $\mathbb{K} : \mathbb{R}^{|\mathcal{I}|} \times \mathbb{R}^{|\mathcal{I}|} \mapsto \mathbb{R}$ over an RKHS $\mathcal{H}$ of a single-layer autoencoder with an activation function $\sigma$ (see \cite{ntk_class_notes} for the NTK derivation of a fully-connected neural network), we reduce the recommendation problem to KRR as follows:

\begin{align} \label{inf_ae_optimization}
\begin{gathered}
    \argmin{[\alpha_j]_{j=1}^{|\mathcal{U}|}} \hspace{0.3cm} \sum_{u \in \mathcal{U}} \ \lVert f(X_u ~|~ \alpha) - X_u \rVert_2^2 + \lambda \cdot \lVert f \rVert_{\mathcal{H}}^2 \\
    \text{s.t.} \hspace{0.3cm} f(X_u ~|~ \alpha) = \sum_{j=1}^{|\mathcal{U}|} \alpha_j \cdot \mathbb{K}(X_u, X_{u_j})
    \hspace{0.2cm} ; \hspace{0.2cm}
    \mathbb{K}(X_u, X_v) = \Tilde{\sigma}(X_u^T X_v) + \Tilde{\sigma'}(X_u^T X_v) \cdot X_u^T X_v
\end{gathered}
\end{align}

Where $\lambda$ is a fixed regularization hyper-parameter, $\alpha \coloneqq [ \alpha_1 ; \alpha_2 \cdots ; \alpha_{|\mathcal{U}|} ] \in \mathbb{R}^{|\mathcal{U}| \times |\mathcal{I}|}$ are the set of dual parameters we are interested in estimating, $\Tilde{\sigma}$ represents the dual activation of $\sigma$ \cite{dual_activation}, and $\Tilde{\sigma'}$ represents its derivative. Defining a gramian matrix $K \in \mathbb{R}^{|\mathcal{U}|\times|\mathcal{U}|}$ where each element can intuitively be seen as the \emph{similarity} of two users, \ie, $K_{i, j} \coloneqq \mathbb{K}(X_{u_i}, X_{u_j})$, the optimization problem defined in \cref{inf_ae_optimization} has a closed-form solution given by $\hat{\alpha} = (K + \lambda I)^{-1} \cdot X$. We can subsequently perform inference for any novel user as follows: $\hat{\text{P}}(\cdot ~|~ u, \mathcal{D}) = \text{softmax}(f(X_u ~|~ \hat{\alpha}))$. We also provide \model's training and inference pseudo-codes in \cref{appendix:pseuocodes}, \cref{alg:inf_ae_trian,alg:inf_ae_predict}.


\paragraph{Scalability.} We carefully examine the computational cost of \model's training and inference. Starting with training, \model has the following computationally-expensive steps: (1) computing the gramian matrix $K$; and (2) performing its inversion. The overall training time complexity thus comes out to be $\mathcal{O}(|\mathcal{U}|^2 \cdot |\mathcal{I}| + |\mathcal{U}|^{2.376})$ if we use the Coppersmith-Winograd algorithm \cite{matrix_inverse} for matrix inversion, whereas the memory complexity is $\mathcal{O}(|\mathcal{U}| \cdot |\mathcal{I}| + |\mathcal{U}|^2)$ for storing the data matrix $X$ and the gramian matrix $K$. As for inference for a single user, both the time and memory requirements are $\mathcal{O}(|\mathcal{U}| \cdot |\mathcal{I}|)$. Observing these computational complexities, we note that \model can be difficult to scale-up to larger datasets na\"ively. To this effect, we address \model's scalability challenges in \sampler (\cref{sec:distillation}), by leveraging a simple observation from support vector machines: not all datapoints (users in our case) are important for model learning. Additionally, in practice, we can perform all of these matrix operations on accelerators like GPU/TPUs and achieve a much higher throughput.

\section{\sampler} \label{sec:distillation}
\paragraph{Motivation.} 
Representative sampling of large datasets has numerous downstream applications. Consequently, in this section we develop \sampler: a data distillation framework to \emph{synthesize} small, high-fidelity data summaries for collaborative filtering (CF) data. We design \sampler with the following rationales: (1) mitigating the scalability challenges in \model by training it only on the terse data summaries generated by \sampler; (2) improving the sample complexity of existing, finite-width recommendation models; (3) addressing the privacy risks of releasing user feedback data by releasing their synthetic data summaries instead; and (4) abating the downstream CO$_2$ emissions of frequent, large-scale recommendation model training by estimating their parameters only on much smaller data summaries synthesized by \sampler. 

\paragraph{Challenges.} Existing work in data distillation has focused on continuous domain data such as images \cite{kip, kip_conv, zhao_dc, zhao_dsa}, and are not directly applicable to heterogeneous and semi-structured domains such as recommender systems and graphs. This problem is further exacerbated since the data for these tasks is severely sparse. For example, in recommendation scenarios, a vast majority of users interact with very few items \cite{pfastre}. Likewise, the nodes in a number of graph-based datasets tend to have connections with very small set of nodes \cite{cold_brew}. We will later show how our \sampler framework is elegantly able to circumvent both these issues while being accurate, and scalable to large datasets.

\paragraph{Methodology.} Given a recommendation dataset $\mathcal{D}$, we aim to synthesize a smaller, support dataset $\mathcal{D}^s$ that can match the performance of recommendation models $\phi : \mathcal{U} \times \mathcal{I} \mapsto \mathbb{R}$ when trained on $\mathcal{D}$ versus $\mathcal{D}^s$. We take inspiration from \cite{kip}, which is also a data distillation technique albeit for images. Formally, given a recommendation model $\phi$, a held-out validation set $\mathcal{D}^{\text{val}}$, and a differentiable loss function $l : \mathbb{R} \times \mathbb{R} \mapsto \mathbb{R}$ that measures the correctness of a prediction with the actual user-item relevance, the data distillation task can be viewed as the following bilevel optimization problem:

\begin{equation} \label{eqn:data_optimization}
    \argmin{\mathcal{D}^s} \hspace{0.3cm} \expectation{(u, i, r) \sim \mathcal{D}^{\text{val}}}{l\left(\phi^*_{\mathcal{D}^s}(u, i), r\right)}
    \hspace{0.3cm} ; \hspace{0.3cm}
    \text{s.t.} \hspace{0.3cm} \phi^*_{\mathcal{D}^s} \coloneqq \argmin{\phi} \hspace{0.3cm} \expectation{(u, i, r) \sim \mathcal{D}^s}{l(\phi(u, i), r)}
\end{equation}

This optimization problem has an \emph{outer loop} which searches for the most informative support dataset $\mathcal{D}^s$ given a fixed recommendation model $\phi^*$, whereas the \emph{inner loop} aims to find the optimal recommendation model for a fixed support set.
In \sampler, we use \model as the model of choice at each step of the inner loop for two reasons: (1) as outlined in \cref{sec:inf_ae}, \model has a closed-form solution with a single hyper-parameter $\lambda$, making the inner-loop extremely efficient; and (2) due to the infinite-width correspondence \cite{ntk, finite_vs_infinite_3}, \model is equivalent to training an infinitely-wide autoencoder on $\mathcal{D}^s$, thereby not compromising on performance.

For the outer loop, we focus only on synthesizing \emph{fake users} (given a fixed user budget $\mu$) through a learnable matrix $X^s \in \mathbb{R}^{\mu \times |\mathcal{I}|}$ which stores the interactions for each fake user in the support dataset. However, to handle the discrete nature of the recommendation problem, instead of directly optimizing for $X^s$, \sampler instead learns a \emph{continuous prior} for each user-item pair, denoting the importance of sampling that interaction in our support set (similar to the notion of propensity \cite{sachdeva_kdd20, rec_as_treatments}). We then sample $\hat{X^s} \sim X^s$ to get our final, discrete support set.

Instead of keeping this sampling operation post-hoc, \ie, after solving for the optimization problem in \cref{eqn:data_optimization}; we perform differentiable Gumbel-sampling \cite{gumbel} on each row (user) in $X^s$ at every optimization step, thereby ensuring search only over sparse and discrete support sets. A notable property of recommendation data is that each user can interact with a variable number of items (this distribution is typically long-tailed due to Zipf's law~\cite{mlperf-dlrm}). We circumvent this dynamic user sparsity issue by taking multiple Gumbel-samples for each user, with replacement. This implicitly gives \sampler the freedom to control the user and item popularity distributions in the generated data summary $\hat{X^s}$ by adjusting the entropy in the prior-matrix $X^s$.


Having controlled for the discrete and dynamic nature of recommendation data by the multi-step Gumbel-sampling trick, we further focus on maintaining the sparsity of the synthesized data. To do so, in addition to the outer-loop reconstruction loss, we add an L1-penalty over $\hat{X^s}$ 
to promote and explicitly control its sparsity 
\cite[Chapter~3]{l1_induces_sparsity}. Furthermore, tuning the number of Gumbel samples we take for each fake user, gives us more control over the sparsity in our generated data summary. More formally, the final optimization objective in \sampler can be written as:

\begin{align} \label{eqn:distill_optimization}
\begin{gathered}
    \argmin{X^s} \hspace{0.3cm} \expectation{u \sim \mathcal{U}}{X_u \cdot \log(K_{X_u\hat{X^s}} \cdot \alpha) + (1 - X_u) \cdot \log(1 - K_{X_u\hat{X^s}} \cdot \alpha)} + \lambda_2 \cdot ||\hat{X^s}||_1 \\
    \text{s.t.} \hspace{0.3cm} \alpha = (K_{\hat{X^s}\hat{X^s}} + \lambda I)^{-1} \cdot \hat{X^s}
    \hspace{0.3cm} ; \hspace{0.3cm}
    \hat{X^s} = \sigma \left( \sum_{i=1}^{\gamma} \text{Gumbel}_{\tau}(\text{softmax}(X^s)) \right)
\end{gathered}
\end{align}

Where, $\lambda_2$ represents an appropriate regularization hyper-parameter for minimizing the L1-norm of the sampled support set $\hat{X}^s$, $K_{XY}$ represents the gramian matrix for the NTK of \model over $X$ and $Y$ as inputs, $\tau$ represents the temperature hyper-parameter for Gumbel-sampling, $\gamma$ denotes the number of samples to take for each fake user in $X^s$, and $\sigma$ represents an appropriate activation function which clips all values over $1$ to be exactly $1$, thereby keeping $\hat{X}^s$ binary. We use hard-tanh in our experiments. We also provide \sampler's pseudo-code in \cref{appendix:pseuocodes}, \cref{alg:distill}.

\paragraph{Scalability.} We now analyze the time and memory requirements for optimizing \cref{eqn:distill_optimization}. The inner loop's major component clearly shares the same complexity as \model. However, since the parameters of \model ($\alpha$ in \cref{inf_ae_optimization}) are now being estimated over the much smaller support set $\hat{X}^s$, the time complexity reduces to $\mathcal{O}(\mu^2 \cdot |\mathcal{I}|)$ and memory to $\mathcal{O}(\mu \cdot |\mathcal{I}|)$, where $\mu$ typically only lies in the range of hundreds for competitive performance. On the other hand, for performing multi-step Gumbel-sampling for each synthetic user, the memory complexity of a na\"ive implementation would be $\mathcal{O(\gamma \cdot \mu \cdot |\mathcal{I}|)}$ since AutoGrad stores all intermediary variables for backward computation. However, since the gradient of each Gumbel-sampling step is independent of other sampling steps and can be computed individually, using \texttt{jax.custom\_vjp}, we reduced its memory complexity to $\mathcal{O(\mu \cdot |\mathcal{I}|)}$, adding nothing to the overall inner-loop complexity. 

For the outer loop, we optimize the logistic reconstruction loss using SGD where we randomly sample a batch of $b$ users from $\mathcal{U}$ and update $X^s$ directly. In totality, for an $\xi$ number of outer-loop iterations, the time complexity to run \sampler is $\mathcal{O}(\xi \cdot (\mu^{2} + b + b \cdot \mu) \cdot |\mathcal{I}|)$, and $\mathcal{O}(b \cdot \mu + (\mu + b) \cdot |\mathcal{I}|)$ for memory. In our experiments, we note convergence in only hundreds of outer-loop iterations, making \sampler scalable even for the largest of the publicly available datasets and practically useful.

\section{Experiments} \label{sec:experiments}

\paragraph{Setup.} We use four recommendation datasets with varying sizes and sparsity characteristics. A brief set of data statistics can be found in \cref{appendix:hyper_params}, \cref{data_stats}. 
For each user in the dataset, we randomly split their interaction history into $80/10/10$\% train-test-validation splits. Following recent warnings against unrealistic dense preprocessing of recommender system datasets \cite{sigir20, wsdm22}, we only prune users that have fewer than $3$ interactions to enforce at least one interaction per user in the train, test, and validation sets. No such preprocessing is followed for items. 

\paragraph{Competitor methods \& evaluation metrics.} We compare \model with various baseline and SoTA competitors as surveyed in recent comparative analyses \cite{recsys_sota_quest, making_progress}. More details on their architectures can be found in \cref{appendix:competitors}. We evaluate all models on a variety of pertinent ranking metrics, namely AUC, HitRate (HR@k), Normalized Discounted Cumulative Gain (nDCG@k), and Propensity-scored Precision (PSP@k), each focusing on different components of the algorithm performance. A notable addition to our list of metrics compared to the literature is the PSP metric \cite{pfastre}, which we adapt to the recommendation use case as an indicator of performance on tail items. The exact definition of all of these metrics can be found in \cref{appendix:metrics}.

\input{tables/model_results}

\paragraph{Training details.} We implement both \model and \sampler using \texttt{JAX} \cite{jax} along with the Neural-Tangents package \cite{neural_tangents} for the relevant NTK computations.\footnote{Our implementation for \model is available at \href{https://github.com/noveens/infinite_ae_cf}{\color{blue}{https://github.com/noveens/infinite\_ae\_cf}}}\textsuperscript{,}\footnote{Our implementation for \sampler is available at \href{https://github.com/noveens/distill_cf}{\color{blue}{https://github.com/noveens/distill\_cf}}} We re-use the official implementation of LightGCN,
and implement the remaining competitors ourselves. To ensure a fair comparison, we conduct a 
hyper-parameter search for all competitors on the validation set. More details on the hyper-parameters for \model, \sampler, and all competitors can be found in \cref{appendix:hyper_params}, \cref{tab:hyper_params}. All of our experiments are performed on a single RTX 3090 GPU, with a random-seed initialization of $42$. Additional training details about \sampler can be found in \cref{appendix:more_training_details}.

\paragraph{Does \model outperform existing methods?} We compare the performance of \model with various baseline and competitor methods in \cref{tab:results}. We also include the results of training \model on data synthesized by \sampler with an additional constraint of having a budget of only $\mu = 500$ synthetic users. For the sake of reference, for our largest dataset (Netflix), this equates to a mere $0.1$\% of the total users. There are a few prominent observations from the results in \cref{tab:results}. First, \model significantly outperforms SoTA recommendation algorithms despite having only a single fully-connected layer, and also being much simpler to train and implement. Second, we note that \model trained on just $500$ users generated by \sampler is able to attain $96-105$\% of \model's performance on the full dataset while also outperforming all competitors trained on the \emph{full} dataset. 

\begin{figure}[!t] 
    \includegraphics[width=\linewidth]{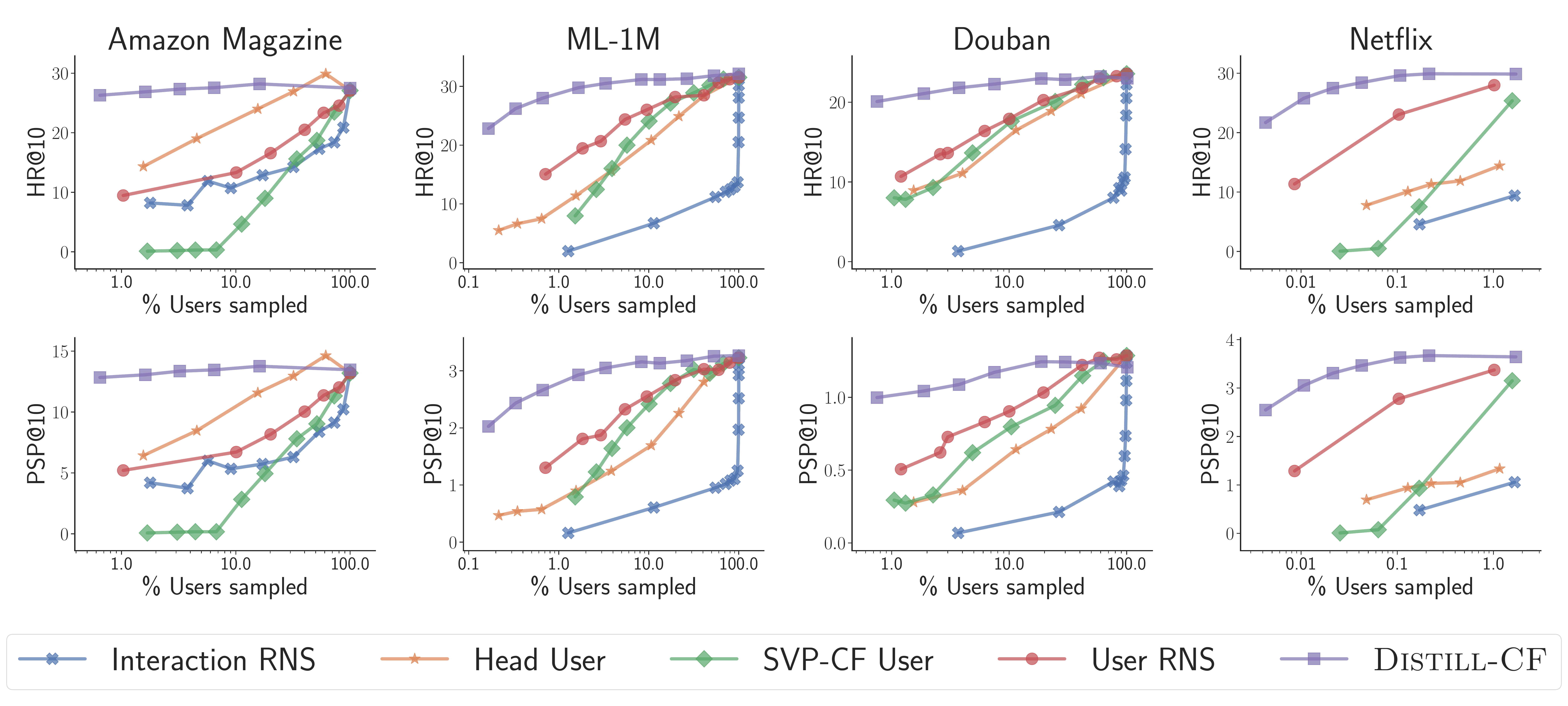}
    \caption{Performance of \model with the amount of users (log-scale) sampled according to different sampling strategies over the HR$@10$ and PSP$@10$ metrics. Results for the Netflix dataset have been clipped due to memory constraints. Results for the remaining metrics can be found in \cref{appendix:extra_experiments}, \cref{fig:inf_ae_sampling_all_metrics}.}
    \label{fig:inf_ae_sampling}
\end{figure}

\paragraph{How sample efficient is \model?} Having noted from \cref{tab:results} that \model is able to outperform all SoTA competitors with as little as $0.1$\% of the original users, we now aim to better understand \model's sample complexity, \ie, the amount of training data \model needs in order to perform accurate recommendation. 
In addition to \sampler, we use the following popular heuristics for down-sampling (more details in \cref{appendix:samplers}): interaction random negative sampling (RNS); user RNS; head user sampling; and a coreset construction technique, SVP-CF user \cite{wsdm22}. We then train \model on sampled data for different sampling budgets, while evaluating on the original test-set. We plot the performance for all datasets computed over the HR@10 and PSP@10 metrics in \cref{fig:inf_ae_sampling}. We observe that while all heuristic sampling strategies tend to be closely bound to the identity line with a slight preference to user RNS, \model when trained on data synthesized by \sampler tends to 
quickly saturate in terms of performance when the user budget is increased, even on the log-scale.
This indicates \sampler's superiority in generating terse data summaries for \model, thereby allowing it to get SoTA performance on the largest datasets with as little as $500$ users.

\begin{figure}[t!] 
    \includegraphics[width=\linewidth]{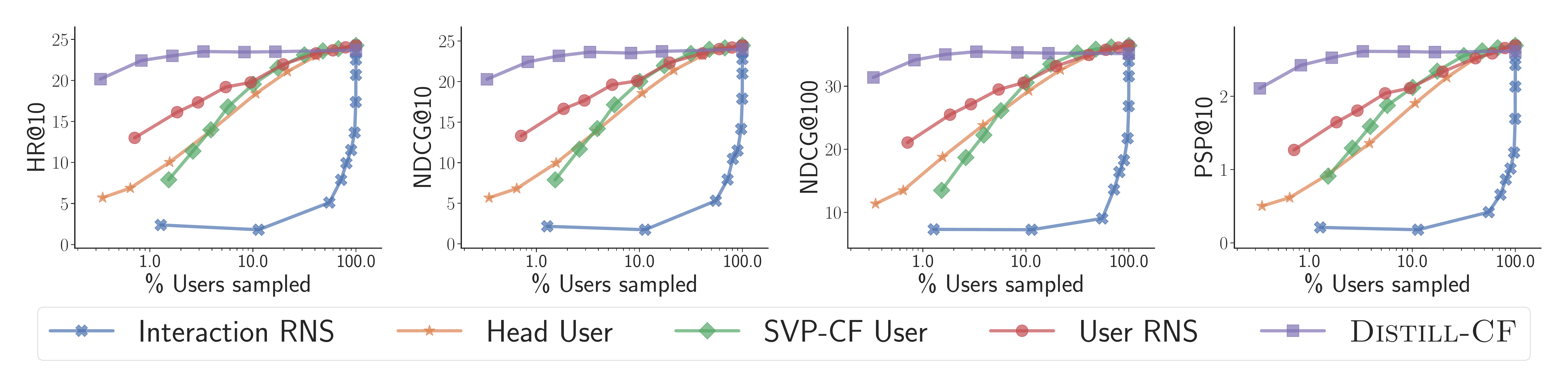}
    \caption{Performance of the EASE model trained on different amounts of users (log-scale) sampled by different samplers on the ML-1M dataset.}
    \label{fig:ease_sampling}
\end{figure}

\paragraph{How transferable are the data summaries synthesized by \sampler?} In order to best evaluate the quality and universality of data summaries synthesized by \sampler, we train and evaluate EASE \cite{ease} 
on data synthesized by \sampler. Note that the inner loop of \sampler still consists of \model, but we nevertheless train and evaluate EASE to test the synthesized data’s universality. 
We re-use the heuristic sampling strategies from the previous experiment for comparison with \sampler. From the results in \cref{fig:ease_sampling}, we observe similar scaling laws as \model's for the heuristic samplers as well as \sampler. 
The semantically similar results for MVAE \cite{mvae} are presented in \cref{appendix:extra_experiments}, \cref{fig:mvae_sampling} for completeness. This behaviour validates the re-usability of data summaries generated by \sampler, because they transfer well to SoTA finite-width models, which were not involved in \sampler's user synthesis optimization.

\begin{figure}[t!] 
    \includegraphics[width=\linewidth]{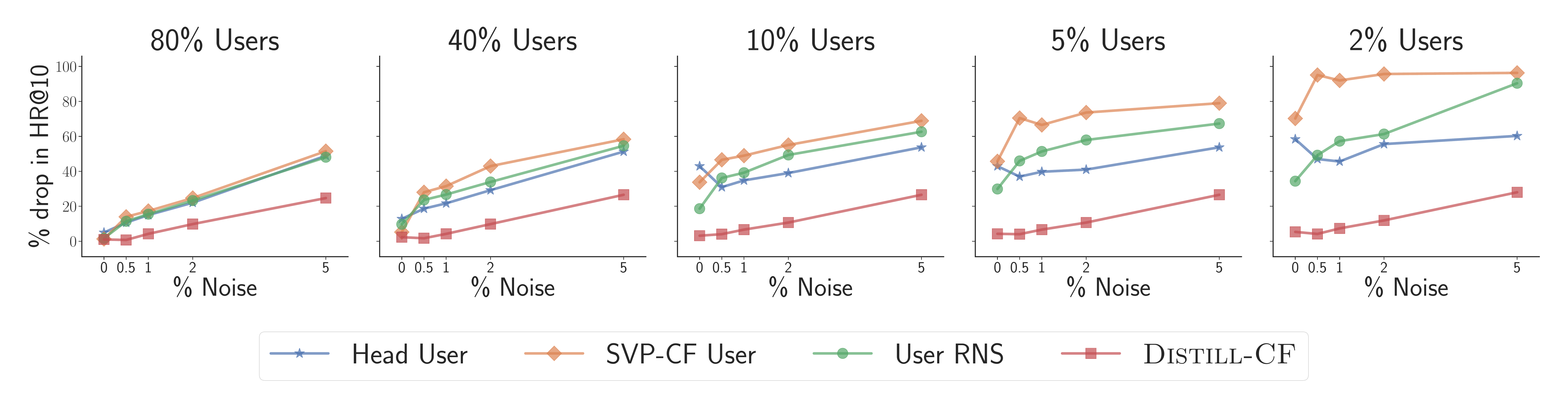}
    \caption{Performance drop of the EASE model trained on data sampled by different sampling strategies when there is varying levels of noise in the data. Performance drop is relative to training on the full, noise-free ML-1M dataset. Results for the remaining metrics can be found in \cref{appendix:extra_experiments}, \cref{fig:denoise_sampler_all_metrics}.}
    \label{fig:denoise_sampler}
\end{figure}

\begin{figure}[t!] 
    \includegraphics[width=\linewidth]{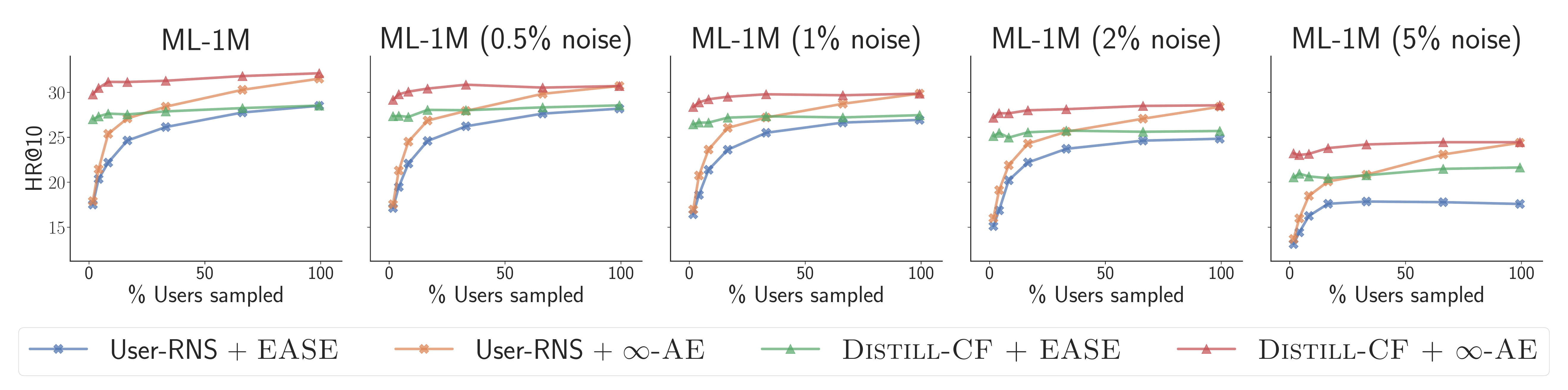}
    \caption{Performance of \model on data sampled by \sampler and User-RNS when there is noise in the data. Results for EASE have been added for reference. All results are on the ML-1M dataset. Results for the remaining metrics can be found in \cref{appendix:extra_experiments}, \cref{fig:denoise_sample_complexity_all_metrics}.}
    \label{fig:denoise_sample_complexity}
\end{figure}

\paragraph{How robust are \sampler and \model to noise?}

User feedback data is often 
noisy due to various biases (see \cite{bias_recsys} for a detailed review). Furthermore, due to the significant  number of logged interactions in these datasets, recommender systems are often trained on down-sampled data in practice.
Despite this, to the best of our knowledge, there is no prior work that explicitly studies the interplay between noise in the data and how sampling it affects downstream model performance.
Consequently, we simulate a simple experiment: we add $x$\% noise in the original train-set $\rightarrow$ sample the noisy training data $\rightarrow$ train recommendation models on the sampled data $\rightarrow$ evaluate their performance on the original, noise-free test-set. For the noise model, we randomly flip $x\%$ of the total number of items in the corpus for each user. In \cref{fig:denoise_sampler}, we compare the drop in HR@10 the EASE model suffers for different sampling strategies when different levels of noise are added to the MovieLens-1M dataset~\cite{movielens}. We make a few 
main
observations: (1) unsurprisingly, sampling noisy data compounds the performance losses of learning algorithms; (2) \sampler has the best noise:sampling:performance trade-off compared to other sampling strategies, with an increasing performance gap relative to other samplers as we inject more noise into the original data; and (3) as we down-sample noisy data more aggressively, head user sampling improves relative to other samplers, simply because these head users are the least affected by our noise injection procedure. \looseness=-1

Furthermore, to better understand \model's denoising capabilities, we repeat the aforementioned noise-injection experiment but now train \model on down-sampled, noisy data. In \cref{fig:denoise_sample_complexity}, we track the change in \model's performance as a function of the number of users sampled, and the amount of noise injected before sampling. We also add the semantically equivalent results for the EASE model for reference. Firstly, we note that the full-data performance-gap between \model and EASE significantly increases when there is more noise in the data, demonstrating \model's robustness to noise, even when its not specifically optimized for it. Furthermore, looking at the $5$\% noise injection scenario, we notice two counter-intuitive observations: (1) training EASE on 
tiny data summaries synthesized by \sampler is better than training it on the full data; and (2) solely looking at data synthesized by \sampler for EASE, we notice the best performance when we have a lower user sampling budget. Both of these observations call for more investigation of a data-centric viewpoint to recommendation, \ie, focusing more on the quality of collected data rather than its quantity.

\paragraph{Applications to continual learning.} 

\begin{wrapfigure}{r}{0.35\textwidth}
  \vspace{-0.8cm}
  \begin{center}
    \includegraphics[width=\linewidth]{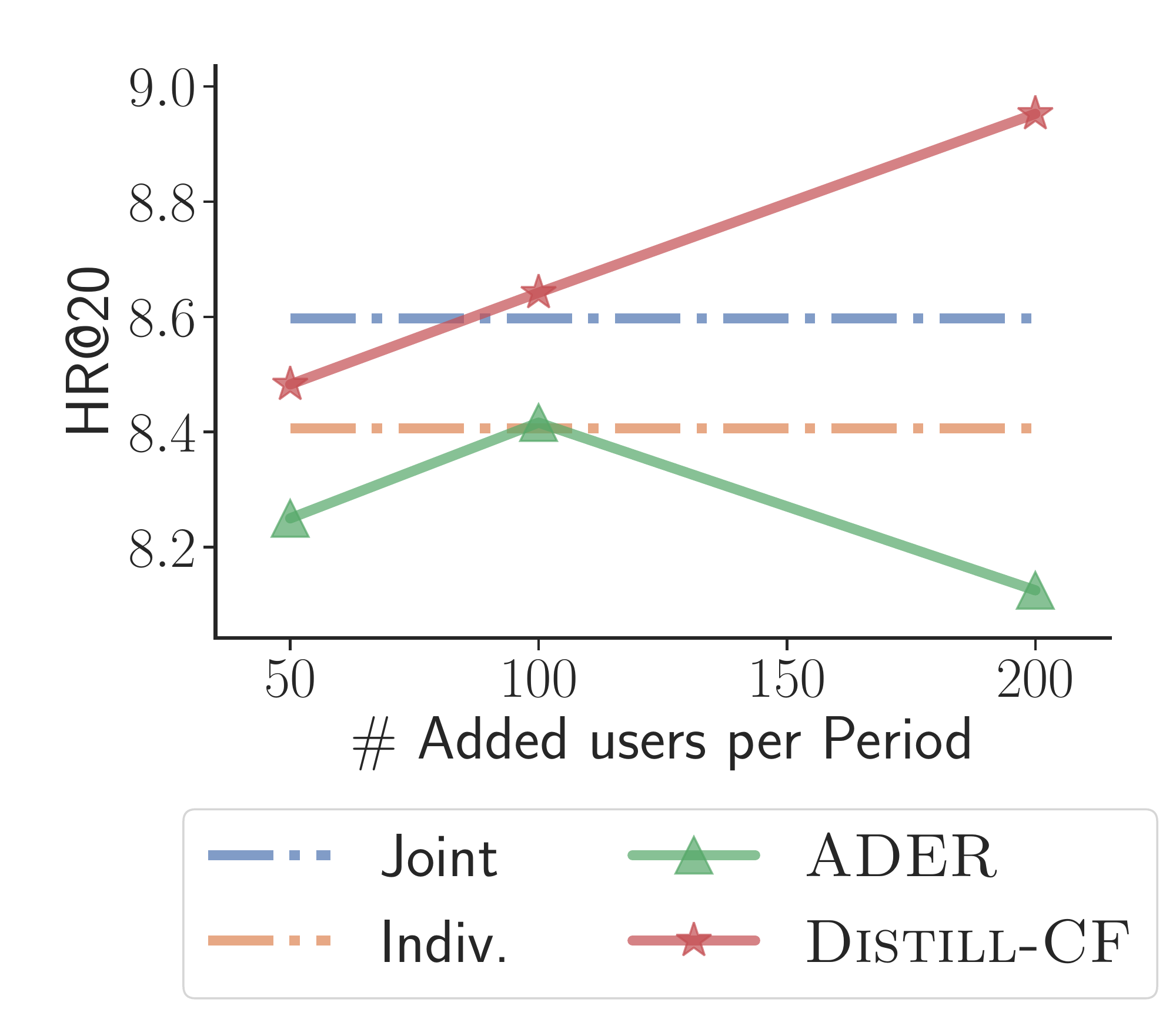}
  \end{center}
  \vspace{-0.2cm}
  \caption{\sampler for continual learning.}
  \label{fig:continual}
  \vspace{-0.3cm}
\end{wrapfigure}

Continual learning (see \cite{continual} for a detailed review) is an important area for recommender systems, because these systems are typically updated at regular intervals. A continual learning scenario involves data that is split into multiple periods, with the predictive task being: given data until the $i^{\text{th}}$ period, maximize algorithm performance for prediction on the $(i+1)^{\text{th}}$ period. \textsc{ADER} \cite{ader} is a SoTA continual learning model for recommender systems, that injects the most informative user sequences from the last period to combat the catastrophic forgetting problem \cite{catast_forgetting}. 
An intuitive application for \sampler is to synthesize succinct data summaries of the last period and inject these instead. 
To compare these approaches, we simulate a continual learning scenario by splitting the MovieLens-1M dataset into $17$ equal sized epochs, and perform experiments with MVAE \cite{mvae} for each period. Note that in \sampler, we still use \model to synthesize data summaries (inner loop).
We also compare with two baselines: (1) Joint: concatenate the data from all periods before the current; and (2) Individual: use the data only from the current period. As we can see from \cref{fig:continual}, \sampler consistently outperforms \textsc{ADER} and the baselines, again demonstrating its ability to generate high-fidelity data summaries.

\section{Conclusion \& Future Work}
In this work, we proposed two complementary ideas: \model, an infinite-width autoencoder for modeling recommendation data, and \sampler for creating tiny, high-fidelity data summaries of massive datasets for subsequent model training. To our knowledge, our work is the first to employ and demonstrate that infinite-width neural networks can beat complicated SoTA models on recommendation tasks. Further, the data summaries synthesized through \sampler outperform generic samplers and demonstrate 
further
performance gains for \model as well as finite-width SoTA models despite being trained on orders of magnitude 
less
data. 

Both our proposed methods are closely linked with one another: \model's closed-loop formulation is especially crucial in the practicality of \sampler, whereas \sampler's ability to distill the entire dataset's knowledge into small summaries helps \model to scale to large datasets. Moreover, the Gumbel sampling trick enables us to adapt data distillation techniques designed for continuous, real-valued, dense domains to heterogeneous, semi-structured, and sparse domains like recommender systems and graphs. We additionally explore the strong denoising effect observed with \sampler, noting that in the case of noisy data, models trained on considerably
less
data synthesized by \sampler perform better than the same model trained on the entire original dataset. \looseness=-1
These observations lead us to contemplate a much larger, 
looming
question: \emph{Is more data what you need for recommendation?} Our results call for further investigation on the data-centric viewpoint of recommendation.

The findings of our paper open up numerous promising research directions. First, building such closed-form, easy-to-implement infinite networks is beneficial for various downstream practical applications like search, sequential recommendation, or CTR prediction. Further, the anonymization achieved by synthesizing fake data summaries is crucial for mitigating the privacy risks associated with confidential or PII datasets. Another direction is analyzing the environmental impact and reduction in carbon footprint as our experiments show that models can achieve similar performance gains when trained on much 
less
data.

\bibliographystyle{plain}
{\small
\bibliography{references}}

\section*{Checklist}


\begin{enumerate}

\item For all authors...
\begin{enumerate}
  \item Do the main claims made in the abstract and introduction accurately reflect the paper's contributions and scope?
    \answerYes{We discuss both \model and \sampler in detail in \cref{sec:inf_ae} and \cref{sec:distillation}, along with validating their performance through experiments in \cref{sec:experiments}}
  \item Did you describe the limitations of your work?
    \answerYes{We talked about the lack of scalability of using \model na\"ively in \cref{sec:inf_ae}}
  \item Did you discuss any potential negative societal impacts of your work?
    \answerNA{}
  \item Have you read the ethics review guidelines and ensured that your paper conforms to them?
    \answerYes{}
\end{enumerate}

\item If you are including theoretical results...
\begin{enumerate}
  \item Did you state the full set of assumptions of all theoretical results?
    \answerNA{}
        \item Did you include complete proofs of all theoretical results?
    \answerNA{}
\end{enumerate}

\item If you ran experiments...
\begin{enumerate}
  \item Did you include the code, data, and instructions needed to reproduce the main experimental results (either in the supplemental material or as a URL)?
    \answerYes{Discussed throughout the setup sub-section in \cref{sec:experiments}, and provided detailed instructions in \cref{appendix:experiments}}
  \item Did you specify all the training details (e.g., data splits, hyperparameters, how they were chosen)?
    \answerYes{Discussed in the setup sub-section in \cref{sec:experiments}, and provided hyper-parameters in \cref{appendix:hyper_params}}
        \item Did you report error bars (e.g., with respect to the random seed after running experiments multiple times)?
    \answerYes{Noted the seed we used in our experiments.}
        \item Did you include the total amount of compute and the type of resources used (e.g., type of GPUs, internal cluster, or cloud provider)?
    \answerYes{Discussed in the training details subsection in \cref{sec:experiments}}
\end{enumerate}

\item If you are using existing assets (e.g., code, data, models) or curating/releasing new assets...
\begin{enumerate}
  \item If your work uses existing assets, did you cite the creators?
    \answerYes{}
  \item Did you mention the license of the assets?
    \answerYes{}
  \item Did you include any new assets either in the supplemental material or as a URL?
    \answerYes{Released our anonymous code as mentioned in \cref{sec:experiments}}
  \item Did you discuss whether and how consent was obtained from people whose data you're using/curating?
    \answerNA{}
  \item Did you discuss whether the data you are using/curating contains personally identifiable information or offensive content?
    \answerNA{}
\end{enumerate}

\item If you used crowdsourcing or conducted research with human subjects...
\begin{enumerate}
  \item Did you include the full text of instructions given to participants and screenshots, if applicable?
    \answerNA{}
  \item Did you describe any potential participant risks, with links to Institutional Review Board (IRB) approvals, if applicable?
    \answerNA{}
  \item Did you include the estimated hourly wage paid to participants and the total amount spent on participant compensation?
    \answerNA{}
\end{enumerate}

\end{enumerate}



\newpage

\appendix

\section{Appendix: Pseudo-codes} \label{appendix:pseuocodes}

\begin{figure*}[h] \centering
\begin{minipage}{0.49\textwidth}
    \centering
    \input{algorithms/inf_ae_train}
\end{minipage} \hfill
\begin{minipage}{0.49\textwidth}
    \centering
    \input{algorithms/inf_ae_predict}
\end{minipage}
\end{figure*} 

\input{algorithms/distillation}

\section{Appendix: Experiments} \label{appendix:experiments}

\subsection{Baselines \& Competitor Methods} \label{appendix:competitors}

We provide a high-level overview of the competitor models used in our experiments:

\begin{itemize}[leftmargin=.3in]
    \item \textbf{PopRec:} This implicit-feedback baseline simply recommends the most \emph{popular} items in the catalog irrespective of the user. Popularity of an item is estimated by their empirical frequency in the logged train-set.
    
    \item \textbf{Bias-only:} This baseline learns scalar user and item biases for all users and item in the train-set, optimized by solving a least-squares regression problem between the predicted and observed relevance. More formally, given a user $u$ and an item $i$, the relevance is predicted as $\hat{r}_{u, i} = \alpha + \beta_u + \beta_i$, where $\alpha \in \mathbb{R}$ is a global offset bias, and $\beta_u, \beta_i \in \mathbb{R}$ are the user and item specific biases respectively. This model doesn't consider any cross user-item interactions, and hence lacks expressivity.
    
    \item \textbf{MF:} Building on top of the bias-only baseline, the Matrix Factorization algorithm tries to represent the users and items in a shared latent space, modeling their relevance by the dot-product of their representations. More formally, given a user $u$ and an item $i$, the relevance is predicted as $\hat{r}_{u, i} = \alpha + \beta_u + \beta_i + (\gamma_u \cdot \gamma_i)$, where $\alpha, \beta_u, \beta_i$ are global, user, and item biases respectively, and $\gamma_u, \gamma_i \in \mathbb{R}^d$ represent the learned user and item representations. The biases and latent representations in this model are estimated by optimizing for the Bayesian Personalized Ranking (BPR) loss \cite{bpr}.
    
    \item \textbf{NeuMF \cite{neural_mf}:} As a neural extension to MF, Neural Matrix Factorization aims to replace the linear cross-interaction between the user and item representations with an arbitrarily complex, non-linear neural network. More formally, given a user $u$ and an item $i$, the relevance is predicted as $\hat{r}_{u, i} = \alpha + \beta_u + \beta_i + \phi(\gamma_u, \gamma_i)$, where $\alpha, \beta_u, \beta_i$ are global, user, and item biases respectively, $\gamma_u, \gamma_i \in \mathbb{R}^d$ represent the learned user and item representations, and $\phi : \mathbb{R}^d \times \mathbb{R}^d \mapsto \mathbb{R}$ is a neural network. The parameters for this model are again optimized using the BPR loss.
    
    \item \textbf{MVAE \cite{mvae}:} This method proposed using variational auto-encoders for the task of collaborative filtering. Their main contribution was to provide a principled approach to perform variational inference for the task of collaborative filtering. 
    
    \item \textbf{LightGCN \cite{light_gcn}:} This simplistic Graph Convolution Network (GCN) framework removes all the steps in a typical GCN \cite{gcn}, only keeping a linear neighbourhood aggregation step. This \emph{light} model demonstrated promising results for the collaborative filtering scenario, despite its simple architecture. We use the official public implementation\footnote{\href{https://github.com/gusye1234/LightGCN-PyTorch}{\color{blue}{https://github.com/gusye1234/LightGCN-PyTorch}}} for our experiments.
    
    \item \textbf{EASE \cite{ease}:} This linear model proposed doing ordinary least squares regression by estimating an item-item similarity matrix, that can be viewed as a zero-depth auto-encoder. Performing regression gives them the benefit of having a closed-form solution. Despite its simplicity, EASE has been shown to out-perform most of the deep non-linear neural networks for the task of collaborative filtering.
\end{itemize}

\input{tables/data_stats}

\subsection{Sampling strategies} \label{appendix:samplers}

Given a recommendation dataset $\mathcal{D} \coloneqq \left\{ (\text{user}_i, \text{item}_i, \text{relevance}_i) \right\}_{i=1}^{n}$ consisting of $n$ user-item interactions defined over a set of users $\mathcal{U}$, set of items $\mathcal{I}$, and operating over a binary relevance score (implicit feedback); we aim to make a $p$\% sub-sample of $\mathcal{D}$, defined in terms of number of interactions. Below are the different sampling strategies we used in comparison with \sampler:

\begin{itemize}[leftmargin=.3in]
    \item \textbf{Interaction-RNS:} Randomly sample $p$\% interactions from $\mathcal{D}$.
    
    \item \textbf{User-RNS:} Randomly sample a user $u \sim \mathcal{U}$, and add all of its interactions into the sampled set. Keep repeating this procedure until the size of sampled set is less than $\frac{p \times n}{100}$.
    
    \item \textbf{Head user:} Sample the user $u$ from $\mathcal{U}$ with the most number of interactions, and add all of its interactions into the sampled set. Remove $u$ from $\mathcal{U}$. Keep repeating this procedure until the size of sampled set is less than $\frac{p \times n}{100}$.
    
    \item \textbf{SVP-CF user \cite{wsdm22}:} This coreset mining technique proceeds by first training a proxy model on $\mathcal{D}$ for $e$ epochs. SVP-CF then modifies the forgetting events approach \cite{forgetting_events}, and counts the inverse AUC for each user in $\mathcal{U}$, averaged over all $e$ epochs. Just like head-user sampling, we now iterate over users in the order of their forgetting count, and keep sampling users until we exceed our sampling budget of $\frac{p \times n}{100}$ interactions. We use the bias-only model as the proxy.
\end{itemize}

\subsection{Data statistics \& hyper-parameter configurations} \label{appendix:hyper_params}

We present a brief summary of statistics of the datasets used in our experiments in \cref{data_stats}, and list all the hyper-parameter configurations tried for \model, \sampler, and other baselines in \cref{tab:hyper_params}.

\input{tables/hyper_params}

\subsection{Additional training details} \label{appendix:more_training_details}

We now discuss additional training details about \sampler that could not be presented in the main text due to space constraints. Firstly, we make use of a validation-set, and evaluate the performance of the \model model trained in \sampler's inner-loop to perform hyper-parameter search, as well as early exit. Note that we don't validate the inner-loop's $\lambda$ at every outer-loop iteration, but keep changing it on-the-fly at each validation cycle. We notice this trick gives us a much faster convergence compared to keeping $\lambda$ fixed for the entire distillation procedure, and validating for it like other hyper-parameters.

We also discuss the Gumbel sampling procedure described in \cref{eqn:distill_optimization}, in more detail. Given the sampling prior matrix $X^s$, that intuitively denotes the importance of sampling a specific user-item interaction, we intend to sample $\hat{X}^s$ which will finally be used for downstream model applications. Note that for each row (user) in $X^s$, we need multiple, but variable number of samples to conform to the Zipfian law for user and item popularity. This requirement in itself rejects the possibility to use top-K sampling which will sample the same number of items for each row. Furthermore, to keep $\hat{X}^s \sim X^s$ sampling part of the optimization procedure, we need to compute the gradients of the logistic objective in \cref{eqn:distill_optimization} with respect to $X^s$, and hence need the entire process to be differentiable. This requirement prohibits the usage of other popular strategies like Nucleus sampling \cite{nucleus}, which is non-differentiable. To workaround all the requirements, we devise a multi-step Gumbel sampling strategy where for each row (user) we take a fixed number of Gumbel samples ($\gamma$), with replacement, followed by taking a union of all the sampled user-item interactions. Note that the union operation ensures that due to sampling with replacement, if a user-item interaction is sampled multiple times, we sample it only once. Hence, the number of sampled interactions is strictly upper-bounded by $\gamma \times |\mathcal{I}|$. To be precise, the sampling procedure is formalized below:

\begin{align*} 
\begin{gathered}
    \hat{X}^s_{u, i} = \sigma \left[ \sum^{\gamma} \frac{exp(\frac{\log(X^s_{u, i}) + g_{u, i}}{\tau})}{\sum_{j \in \mathcal{I}} exp(\frac{\log(X^s_{u, j}) + g_{u, j}}{\tau})} \right] \hspace{0.3cm} \text{s.t.} \hspace{0.3cm} g_{u, i} \sim \text{Gumbel}(\mu = 0, \beta = 1) \hspace{0.3cm} \forall u \in \mathcal{U}, i \in \mathcal{I}
\end{gathered}
\end{align*}

Where $\sigma$ represents an appropriate function which clamps all values between $0$ and $1$. In our experiments, we use hard-tanh.

\subsection{Evaluation metrics} \label{appendix:metrics}

We now present formal definitions of all the ranking metrics used in this study:

\begin{itemize}[leftmargin=.3in]
    \item \textbf{AUC:} Intuitively defined as a threshold independent classification performance measure, AUC can also be interpreted as the expected probability of a recommender system ranking a positive item over a negative item for any given user. More formally, given a user $u$ from the user set $\mathcal{U}$ with its set of positive interactions $\mathcal{I}_u^+ \subseteq \mathcal{I}$ with a similarly defined set of negative interactions $\mathcal{I}_u^- = \mathcal{I} \backslash \mathcal{I}_u^+$, the AUC for a relevance predictor $\phi(u, i)$ is defined as:
    
    \begin{equation*}
        \text{AUC}(\phi) \coloneqq \expectation{u \sim \mathcal{U}}{\expectation{i^+ \sim \mathcal{I}_u^+}{\expectation{i^- \sim \mathcal{I}_u^-}{\phi(u, i^+) > \phi(u, i^-)}}}
    \end{equation*}
    
    \item \textbf{HitRate (HR@k):} Another name for the recall metric, this metric estimates how many positive items are predicted in a top-k recommendation list. More formally, given recommendation lists $\hat{Y}_u \subseteq \mathcal{I}^K ~~~ \forall u \in \mathcal{U}$ and the set of positive interactions $\mathcal{I}_u^+ \subseteq \mathcal{I} ~~~ \forall u \in \mathcal{U}$:
    
    \begin{equation*}
        \text{HR@k} \coloneqq \expectation{u \sim \mathcal{U}}{\frac{|\hat{Y}_u \cap \mathcal{I}_u^+|}{|\mathcal{I}_u^+|}}
    \end{equation*}
    
    \item \textbf{Normalized Discounted Cumulative Gain (nDCG@k):} Unlike HR@k which gives equal importance to all items in the recommendation list, the nDCG@k metric instead gives a higher importance to items predicted higher in the recommendation list and performs logarithmic discounting further down. More formally, given \emph{sorted} recommendation lists $\hat{Y}_u \subseteq \mathcal{I}^K ~~~ \forall u \in \mathcal{U}$ and the set of positive interactions $\mathcal{I}_u^+ \subseteq \mathcal{I} ~~~ \forall u \in \mathcal{U}$:
    
    \begin{equation*}
        \text{nDCG@k} \coloneqq \expectation{u \sim \mathcal{U}}{\frac{\text{DCG}_u}{\text{IDCG}_u}} ~;~ \text{DCG}_u \coloneqq \sum_{i=1}^{k} \frac{\hat{Y}_u^i \in \mathcal{I}_u^+}{log_2(i+1)} ~;~ \text{IDCG}_u \coloneqq \sum_{i=1}^{|\mathcal{I}_u^+|} \frac{1}{log_2(i+1)}
    \end{equation*}
    
    \item \textbf{Propensity-scored Precision (PSP@k):} Originally introduced in \cite{pfastre} for extreme classification scenarios \cite{parabel, slice, eclare}, the PSP@k metric intuitively accounts for missing labels (items in the case of recommendation) by dividing the true relevance of an item (binary) with a propensity correction term. More formally, given recommendation lists $\hat{Y}_u \subseteq \mathcal{I}^K ~~~ \forall u \in \mathcal{U}$, the set of positive interactions $\mathcal{I}_u^+ \subseteq \mathcal{I} ~~~ \forall u \in \mathcal{U}$, and a propensity model $\phi : \mathcal{I} \mapsto \mathbb{R}$:
    
    \begin{equation*}
        \text{PSP@k} \coloneqq \expectation{u \sim \mathcal{U}}{\frac{\text{uPSP}_u}{\text{mPSP}_u}} ~;~ \text{uPSP}_u \coloneqq \frac{1}{k} \cdot \sum_{i=1}^{k} \frac{\hat{Y}_u^i \in \mathcal{I}_u^+}{\phi(\hat{Y}_u^i)} ~;~ \text{mPSP}_u \coloneqq \sum_{i \in \mathcal{I}_u^+} \frac{1}{\phi(i)}
    \end{equation*}
    
    For $\phi$, we adapt the propensity model proposed in \cite{pfastre} for the case of recommendation as:
    
    \begin{equation*}
        \phi(i) \equiv \expectation{u \sim \mathcal{U}}{\text{P}(r_{u, i} = 1 | r^*_{u, i} = 1)} = \frac{1}{1 + C \cdot e^{-A \cdot log(n_i + B)}} ~;~ C = (logN - 1) \cdot (B+1)^A
    \end{equation*}
    
    Where, $N$ represents the total number of interactions in the dataset, and $n_i$ represents the empirical frequency of item $i$ in the dataset. We use $A = 0.55$ and $B = 1.5$ for our experiments.
\end{itemize}

\subsection{Additional experiments} \label{appendix:extra_experiments}

\begin{figure}[ht!] 
    \centering
    \includegraphics[width=0.72\linewidth]{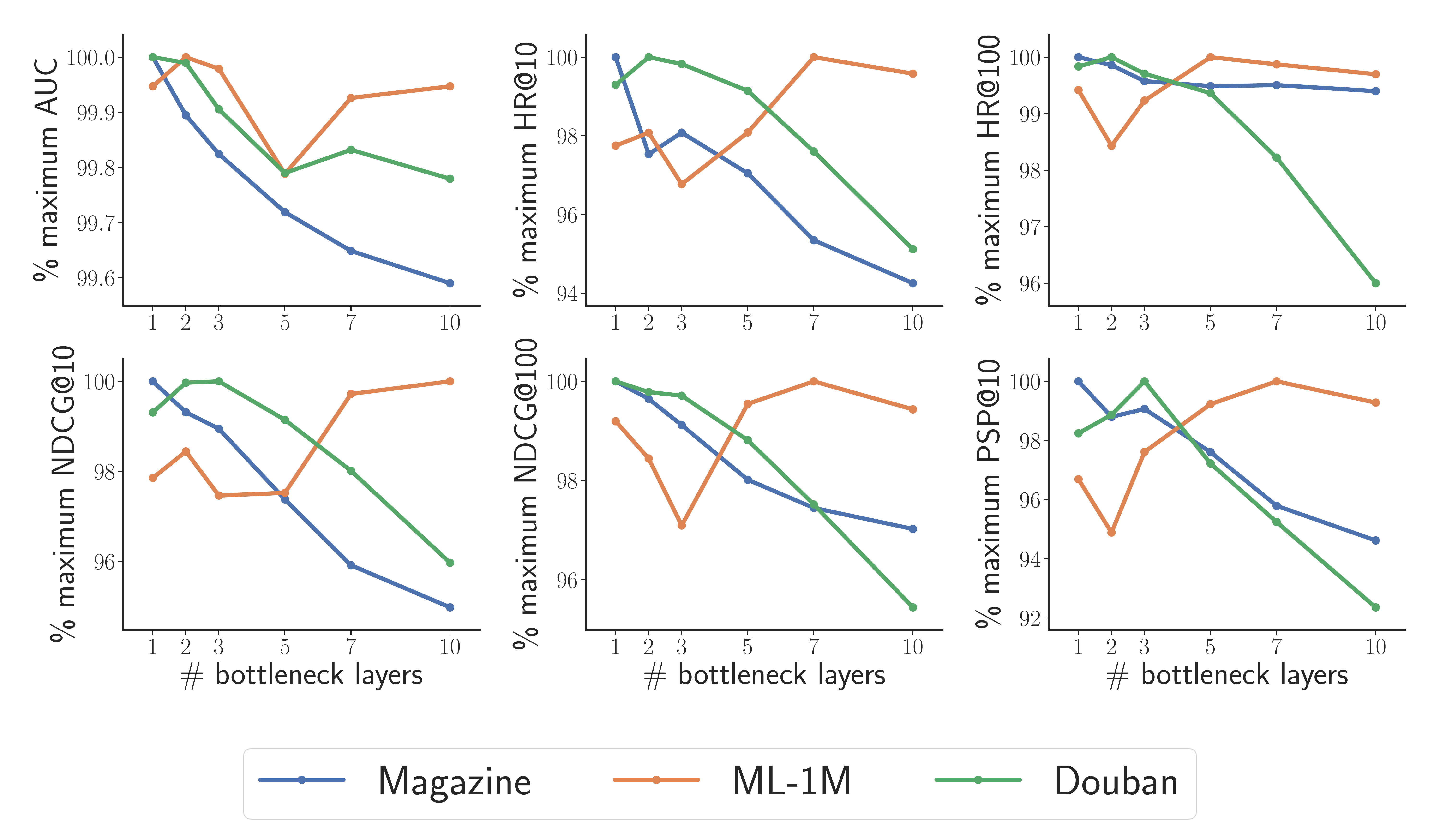}
    \caption{Performance of \model with varying depths. The y-axis represents the normalized metric \ie performance relative to the best depth for a given metric.}
    \label{fig:inf_ae_depth}
\end{figure}

\paragraph{How does depth affect \model?} To better understand the effect of depth on an infinitely-wide auto-encoder's performance for recommendation, we extend \model to multiple layers and note its downstream performance change in \cref{fig:inf_ae_depth}. The prominent observation is that models tend to get \emph{worse} as they get deeper, with generally good performance in the range of $1-2$ layers, which also has been common practical knowledge even for finite-width recommender systems.

\begin{figure}[ht!] 
    \centering
    \includegraphics[width=0.72\linewidth]{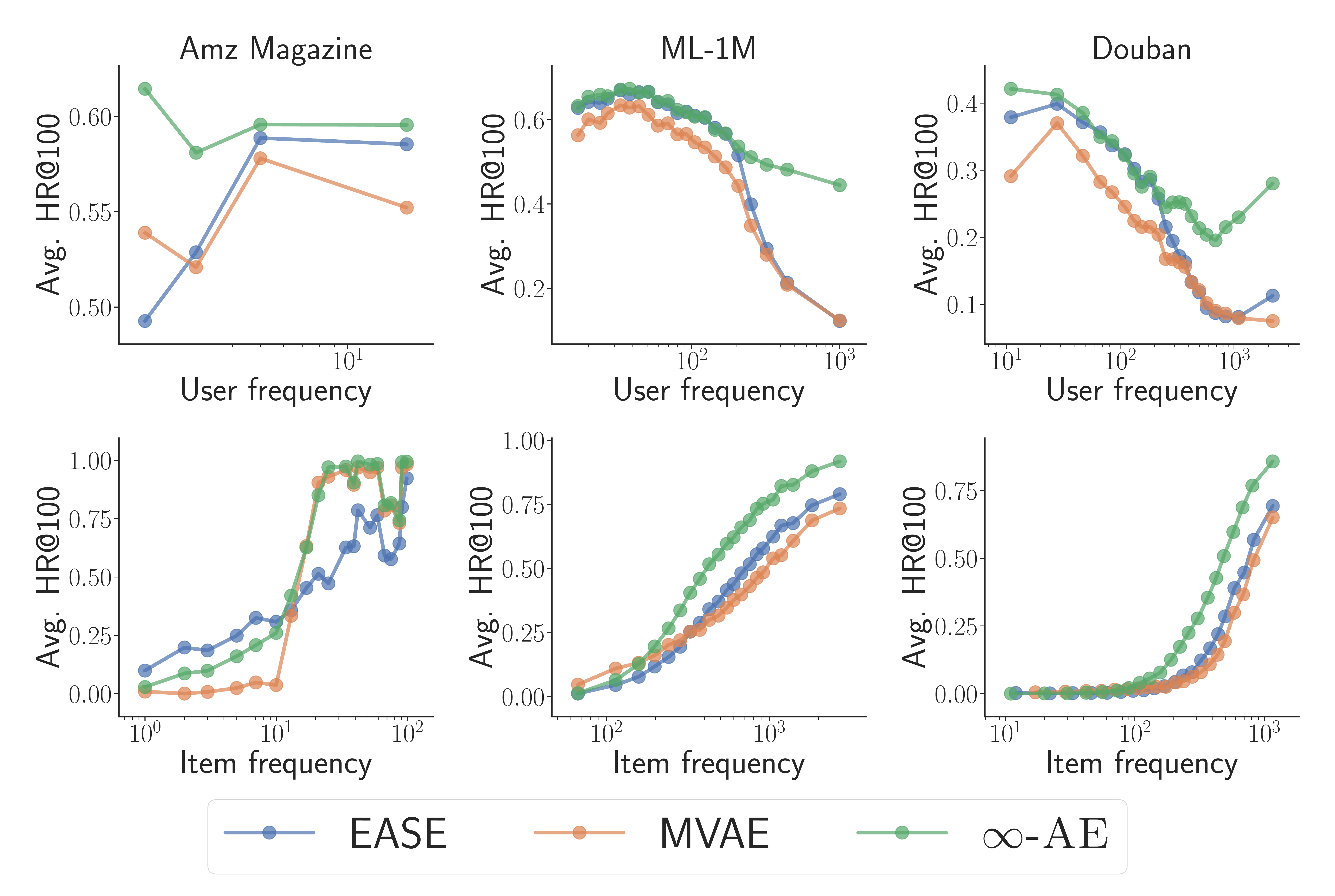}
    \caption{Performance comparison of \model with SoTA finite-width models stratified over the coldness of users and items. The y-axis represents the average HR@100 for users/items in a particular quanta. All user/item bins are equisized.}
    \label{fig:inf_ae_coldness}
\end{figure}

\paragraph{How does \model perform on cold users \& cold items?} Cold-start has been one of the hardest problems in recommender systems --- how to best model users or items that have very little training data available? Even though \model doesn't have any extra modeling for these scenarios, we try to better understand the performance of \model over users' and items' coldness spectrum. In \cref{fig:inf_ae_coldness}, we quantize different users and items based on their coldness (computed by their empirical occurrence in the train-set) into equisized buckets and measure different models' performance only on the binned users or items. We note \model's dominance over other competitors especially over the tail, head-users; and torso, head-items. 

\paragraph{How anonymized is the data synthesized by \sampler?} Having evaluated the fidelity of distills generated using \sampler, we now focus on understanding its anonymity and syntheticity. For the generic data down-sampling case, the algorithm presented in \cite{netflix_deanon} works well to de-anonymize the Netflix prize dataset. The algorithm assumes a complete, non-PII dataset $\mathcal{D}$ along with an incomplete, noisy version of the same dataset $\mathcal{D}'$, but also has the sensitive PII available. We simulate a similar setup but extend to datasets other than Netflix, by following a simple down-sampling and noise addition procedure: given a sampling strategy $s$, down-sample $\mathcal{D}$ and add $x$\% noise by randomly flipping $x$\% of the total items for each user to generate $\mathcal{D}'$. We then use our implementation of the algorithm proposed in \cite{netflix_deanon} to \emph{match} the corresponding users in the original, noise-free dataset $\mathcal{D}$. 

However, if instead of a down-sampled dataset $\mathcal{D}'$, a data distill of $\mathcal{D}$ (using \sampler), let's say $\Tilde{\mathcal{D}}$, is made publicly available. The task of de-anonymization can no longer be carried out by simply \emph{matching} user histories from $\mathcal{D}'$ to $\mathcal{D}$, since $\mathcal{D}$ is no longer available. The only solution now is to \emph{predict} the missing items in $\mathcal{D}'$. Note that this this task is easier than the usual recommendation problem, as the user histories to complete in $\mathcal{D}'$ do exist in some incoherent way in the data distill $\Tilde{\mathcal{D}}$, and is more similar to train-set prediction. To test this out, we formulate a simple experiment: given a data distill $\Tilde{\mathcal{D}}$, an incomplete, noisy subset $\mathcal{D}'$ with PII information, and also hypothetically the number of missing items for each user in $\mathcal{D}'$ --- how accurately can we predict the \emph{exact} set of missing items in $\mathcal{D}'$ using an \model model trained on $\Tilde{\mathcal{D}}$.

\begin{figure}
\centering
\begin{minipage}{.48\textwidth}
  \centering
  \includegraphics[width=\linewidth]{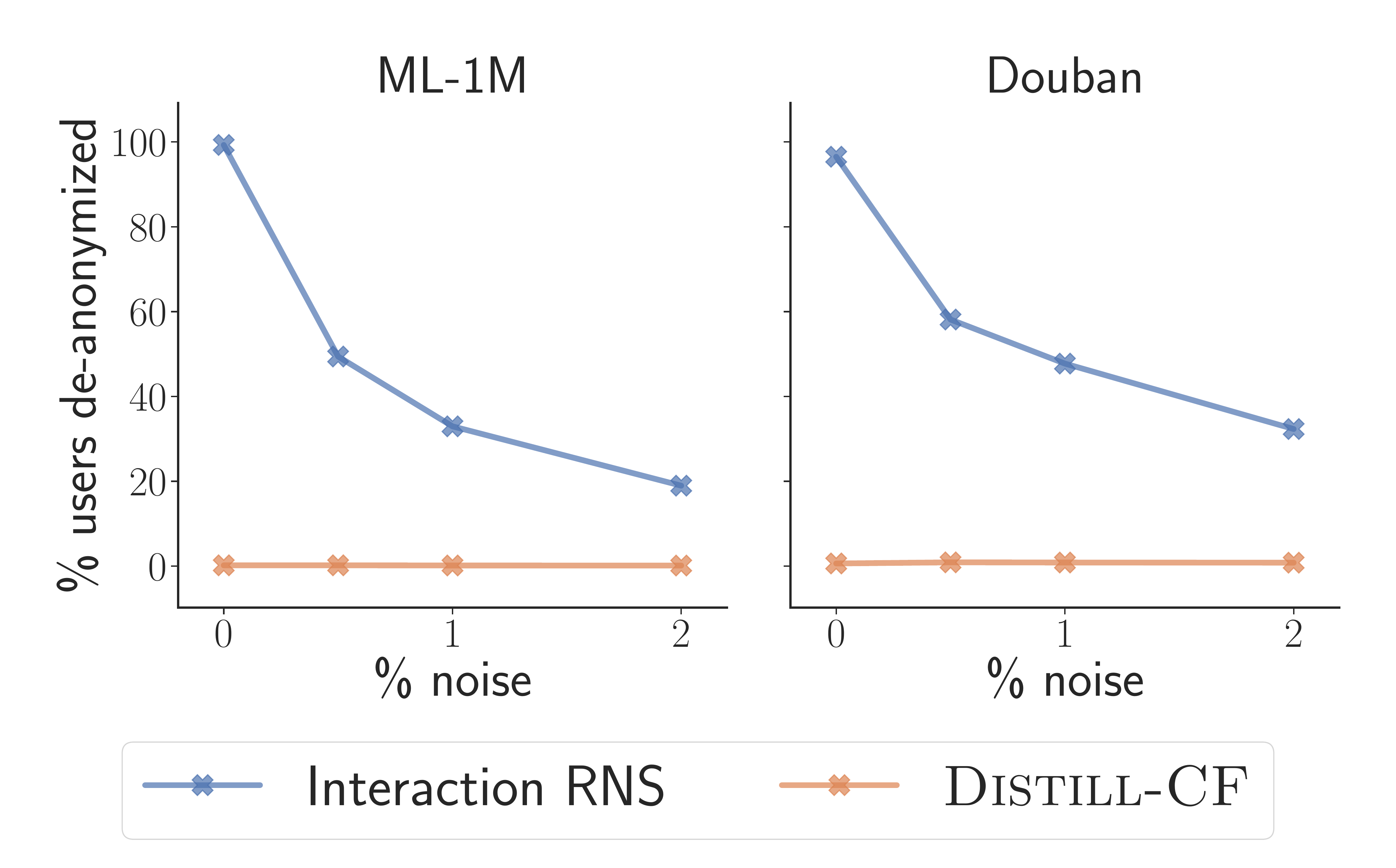}
  \captionof{figure}{Amount of noise added in $\mathcal{D}'$ \vs \% of users de-anonymized.}
  \label{fig:deanon_noise}
\end{minipage} \hfill
\begin{minipage}{.48\textwidth}
  \centering
  \includegraphics[width=\linewidth]{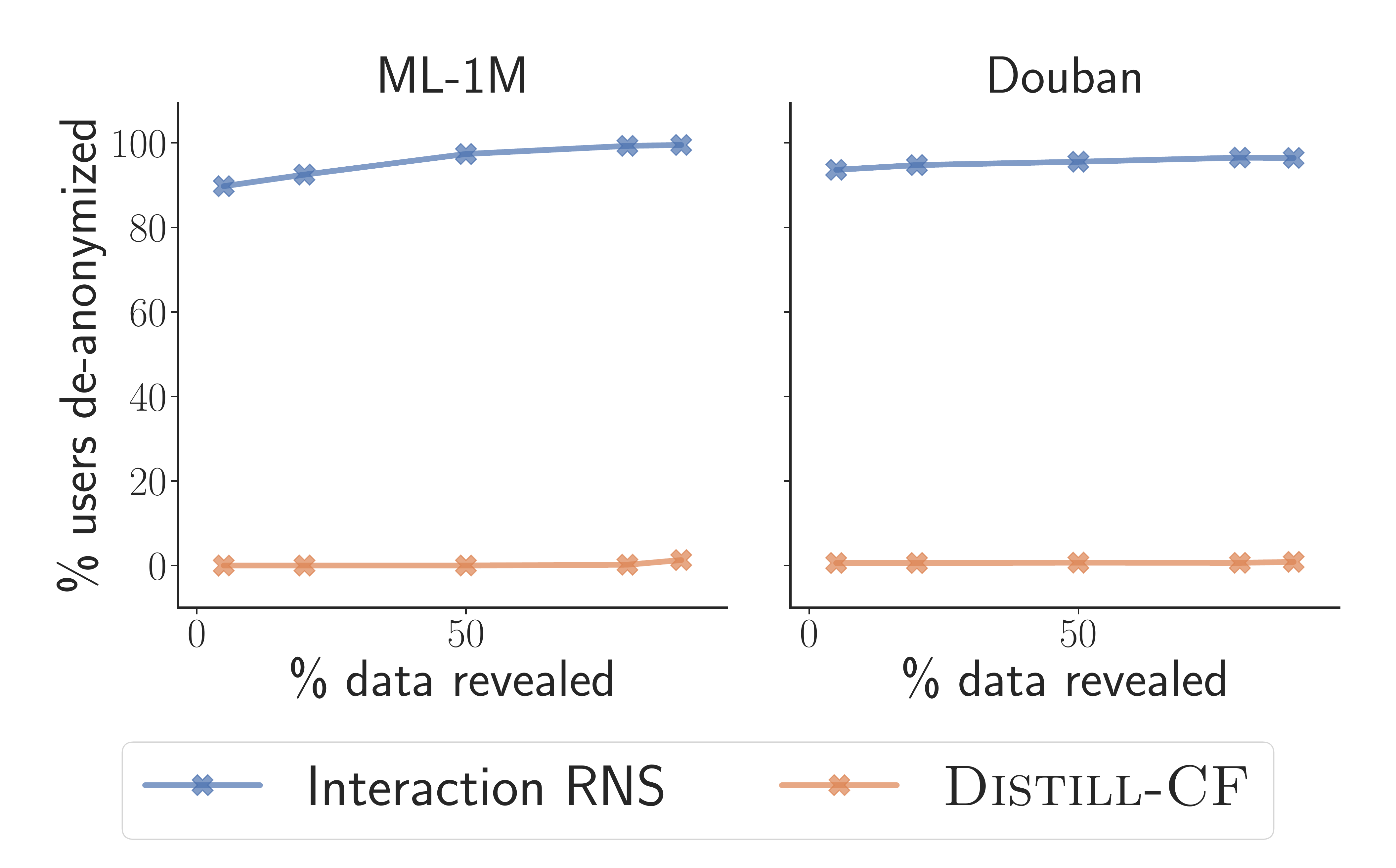}
  \captionof{figure}{Amount of data revealed in $\mathcal{D}'$ \vs \% of users de-anonymized.}
  \label{fig:deanon_data_reveal}
\end{minipage}
\end{figure}

We perform experiments for both the cases of data-sampling and data-distillation. In \cref{fig:deanon_noise}, we measure the \% of users de-anonymized using the aforementioned procedures. We interestingly note no level of de-anonymization with the data-distill, even if there's no noise in $\mathcal{D}'$. We also note the expected observation for the data-sampling case: less users are de-anonymized when there's more noise in $\mathcal{D}'$. In \cref{fig:deanon_data_reveal}, we now control the amount of data revealed in $\mathcal{D}'$. We again note the same observation: even with $90$\% of the correct data from $\mathcal{D}$ revealed in $\mathcal{D}'$ with $0$\% of noise, we still note a very tiny $0.86$\% of user de-anonymization with data-distillation, whereas $96.43$\% with data-sampling for the Douban dataset.

\begin{figure}[ht!]
\centering
\begin{minipage}{.48\textwidth}
  \centering
  \includegraphics[width=0.8\linewidth]{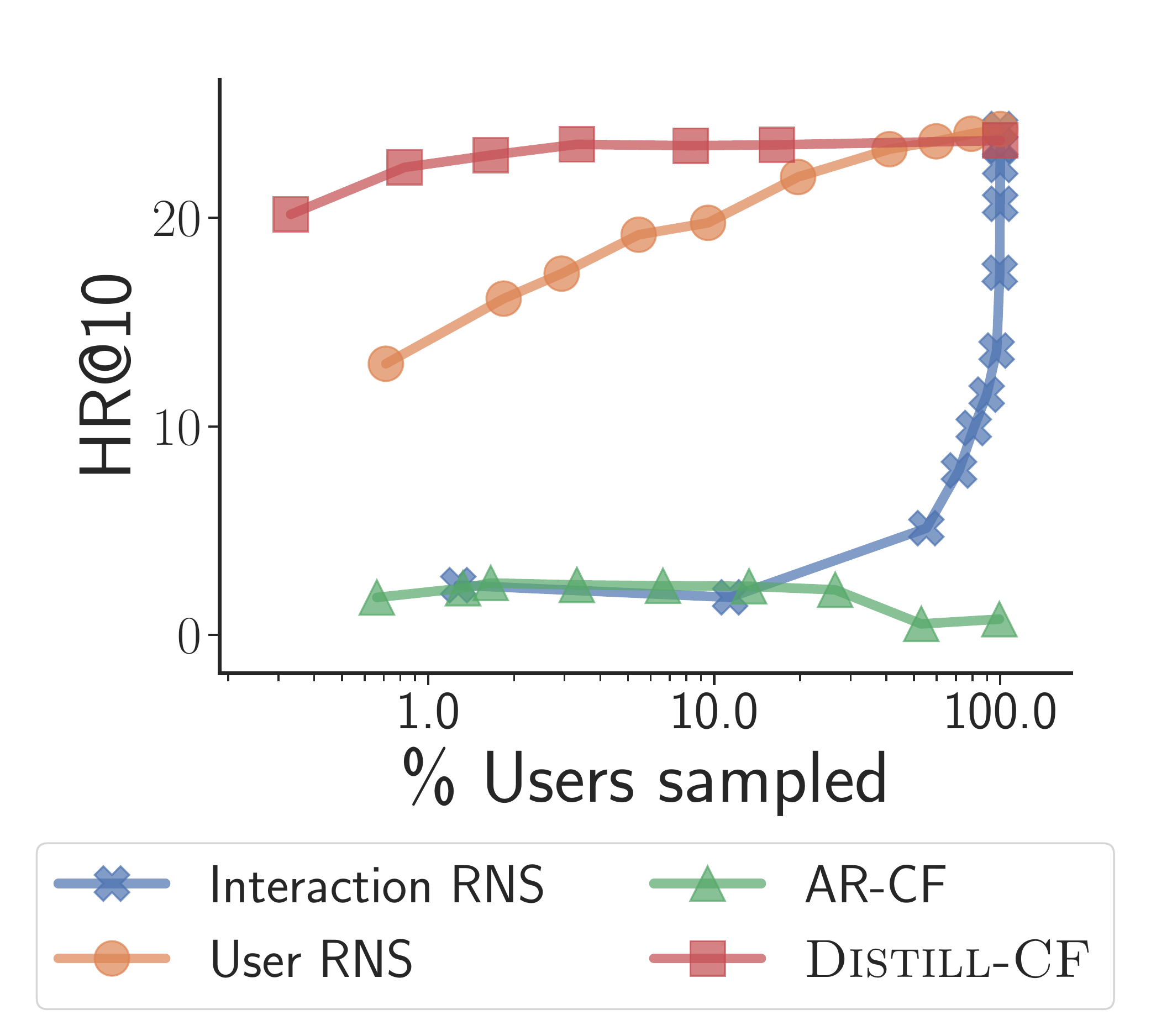}
  \captionof{figure}{Performance of EASE on varying amounts of data sampled/synthesized using various strategies for the MovieLens-1M dataset.}
  \label{fig:arcf}
\end{minipage} \hfill
\begin{minipage}{.48\textwidth}
  \centering
  \includegraphics[width=0.82\linewidth]{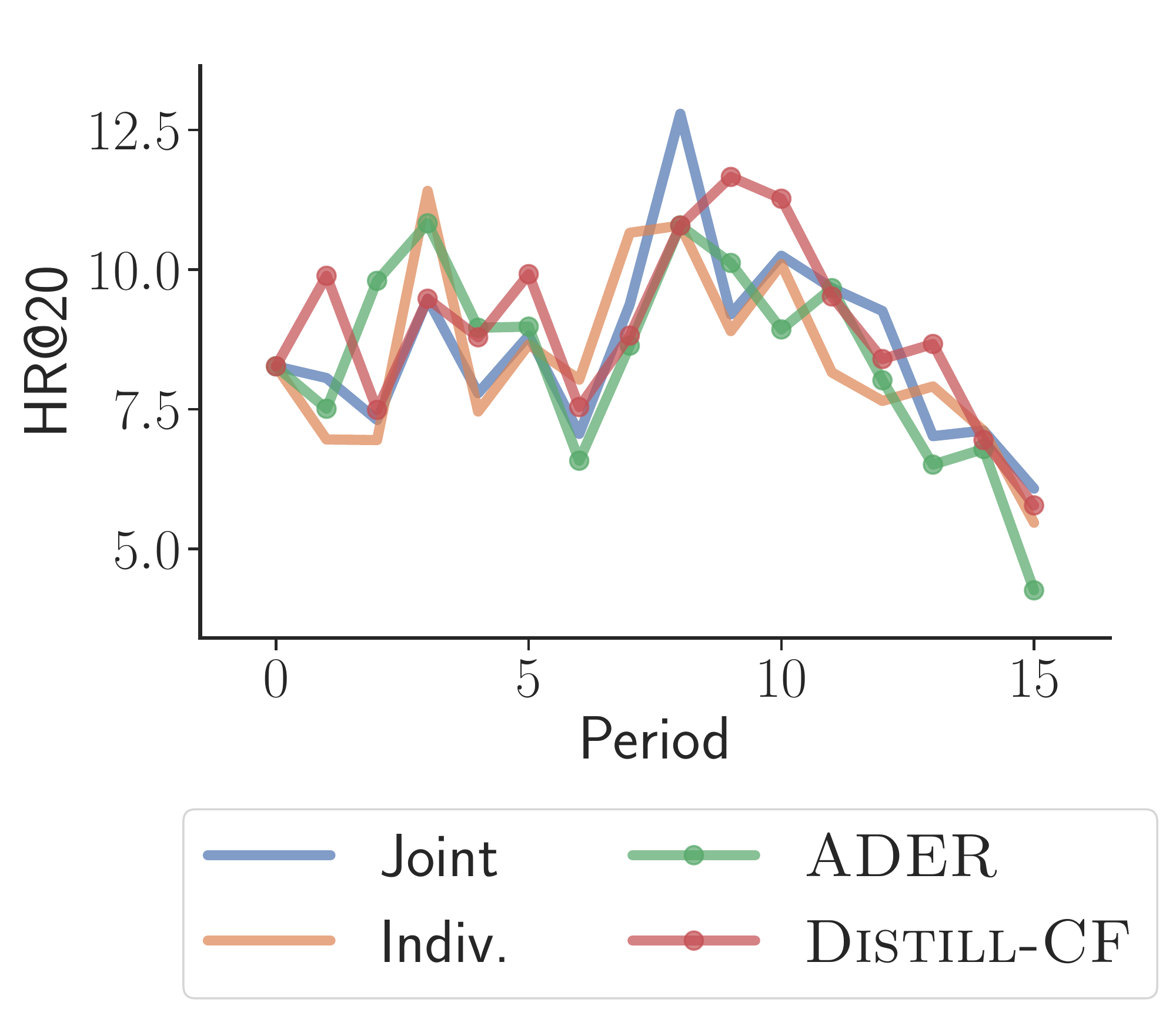}
  \captionof{figure}{Per-period evaluation of the MVAE model on various continual learning strategies as discussed in \cref{sec:experiments}.}
  \label{fig:continual_per_day}
\end{minipage}
\end{figure}

\paragraph{How does \sampler compare to data augmentation approaches?} 
We compare the quality of data synthesized by \sampler with generative models proposed for data augmentation. One such SoTA method is AR-CF \cite{arcf}, which leverages two conditional GANs \cite{cgan} to generate fake users and fake items. For our experiment, we focus only on AR-CF's user generation sub-network and consequently train the EASE \cite{ease} model \emph{only} on these synthesized users, while testing on the original test-set for the MovieLens-1M dataset. We plot the results in \cref{fig:arcf}, comparing the amount of users synthesized according to different strategies and plot the HR@10 of the correspondingly trained model. The results signify that training models only on data synthesized by data augmentation models is impractical, as these users have only been optimized for being \emph{realistic}, whereas the users synthesized by \sampler are optimized to be \emph{informative for model training}. The same observation tends to hold true for the case of images as well \cite{zhao_dc}.

\begin{figure}[ht!] 
    \includegraphics[width=\linewidth]{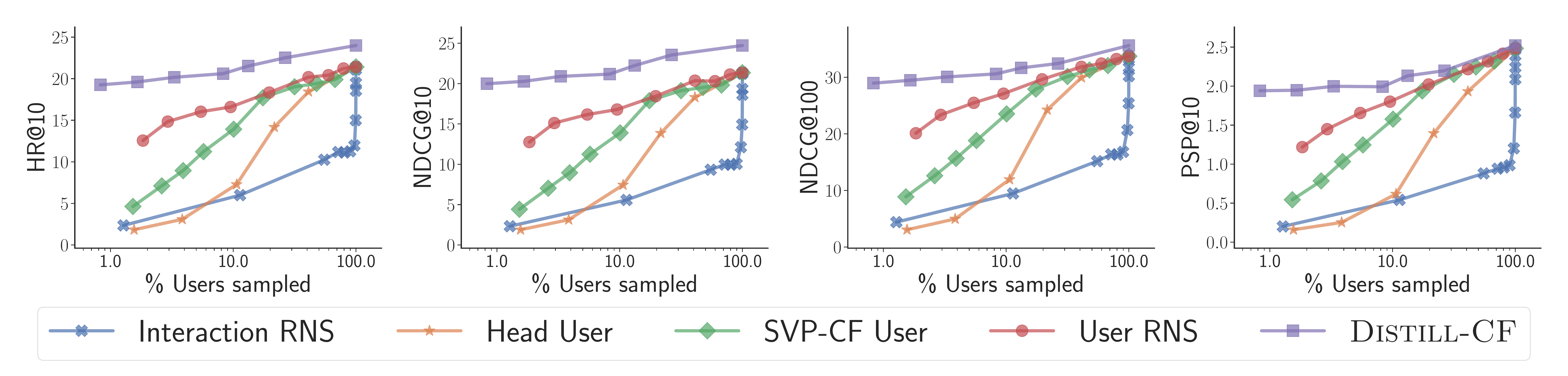}
    \caption{Performance of the MVAE model trained on different amounts of users (log-scale) sampled by different samplers on the ML-1M dataset.}
    \label{fig:mvae_sampling}
\end{figure}

\paragraph{Additional experiments on the generality of data summaries synthesized by \sampler.} Continuing the results in \cref{fig:ease_sampling}, we now train and evaluate MVAE \cite{mvae} on data synthesized by \sampler. Note that the inner loop of \sampler still consists of \model, but we nevertheless train and evaluate MVAE to test the synthesized data's universality. We re-use the heuristic sampling strategies from \cref{fig:ease_sampling} for comparison with \sampler. From the results in \cref{fig:mvae_sampling}, we observe similar scaling laws as EASE for the heuristic samplers as well as \sampler. A notable exception is interaction RNS, that scales like \model. Another interesting observation to note is that training MVAE on the full data performs slightly worse than training MVAE on the same amount of data synthesized by \sampler. This behaviour validates the re-usability of data summaries generated by \sampler, because they transfer well to SoTA finite-width models, which were not involved in \sampler's user synthesis procedure.

\paragraph{Additional experiments on Continual learning.} Continuing the results in \cref{fig:continual}, we plot the results stratified per period for the MovieLens-1M dataset in \cref{fig:continual_per_day}. The results are a little noisy, but we can observe that exemplar data distilled with \sampler has better performance for a majority of the data periods. Note that we use the official public implementation\footnote{\href{https://github.com/doublemul/ADER}{\color{blue}{https://github.com/doublemul/ADER}} available with the MIT license.} of \textsc{ADER}.

\begin{figure}[ht!] 
    \includegraphics[width=\linewidth]{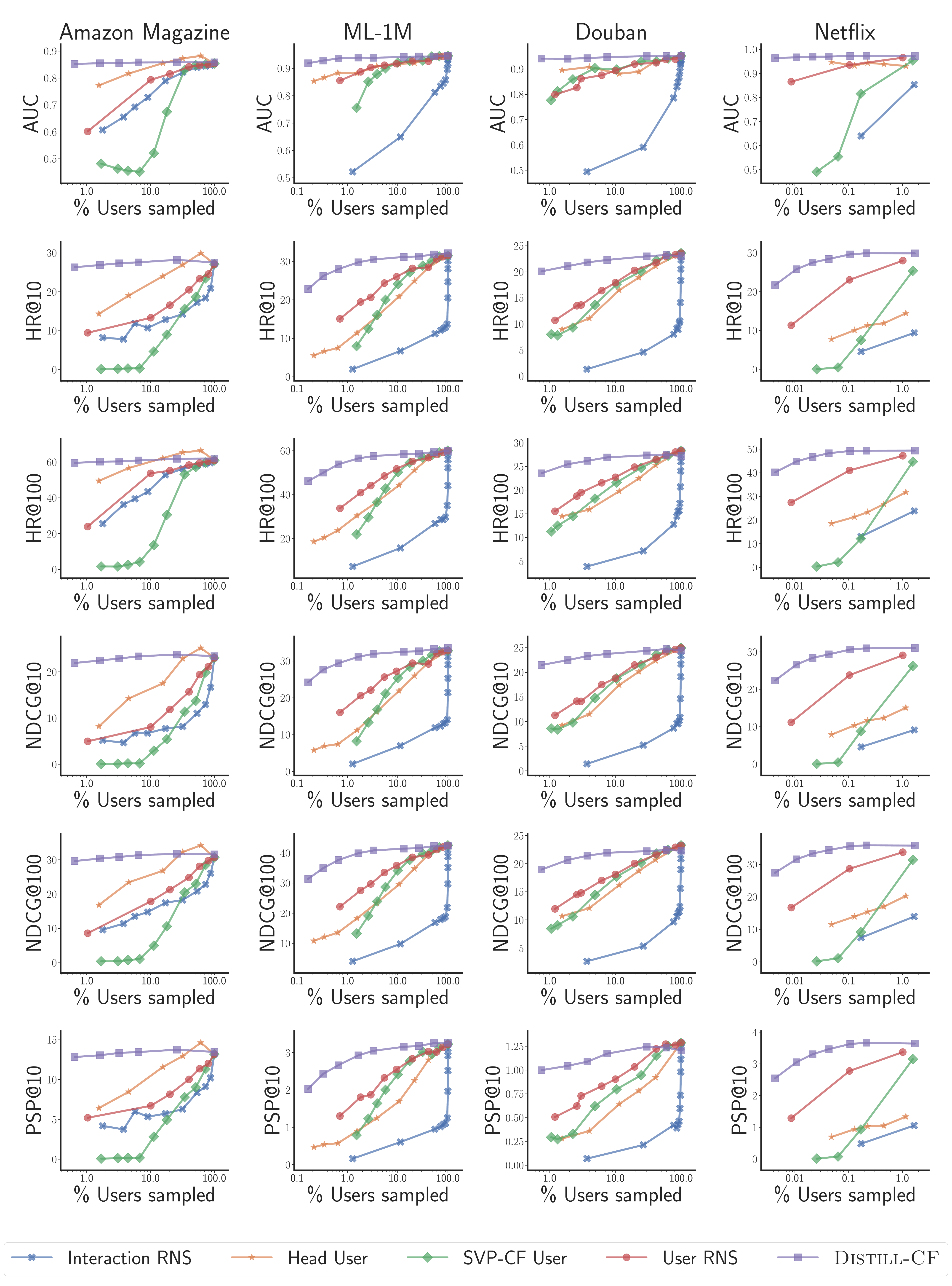}
    \caption{Performance of \model with the amount of users sampled according to different sampling strategies over different metrics. Each column represents a single dataset, and each row represents an evaluation metric.}
    \label{fig:inf_ae_sampling_all_metrics}
\end{figure}

\paragraph{Additional plots for the sample complexity of \model.} In addition to \cref{fig:inf_ae_sampling} in the main paper, we visualize the sample complexity of \model for all datasets and all metrics in \cref{fig:inf_ae_sampling_all_metrics}. We notice similar trends for all metrics across datasets.

\begin{figure}[ht!] 
    \includegraphics[width=\linewidth]{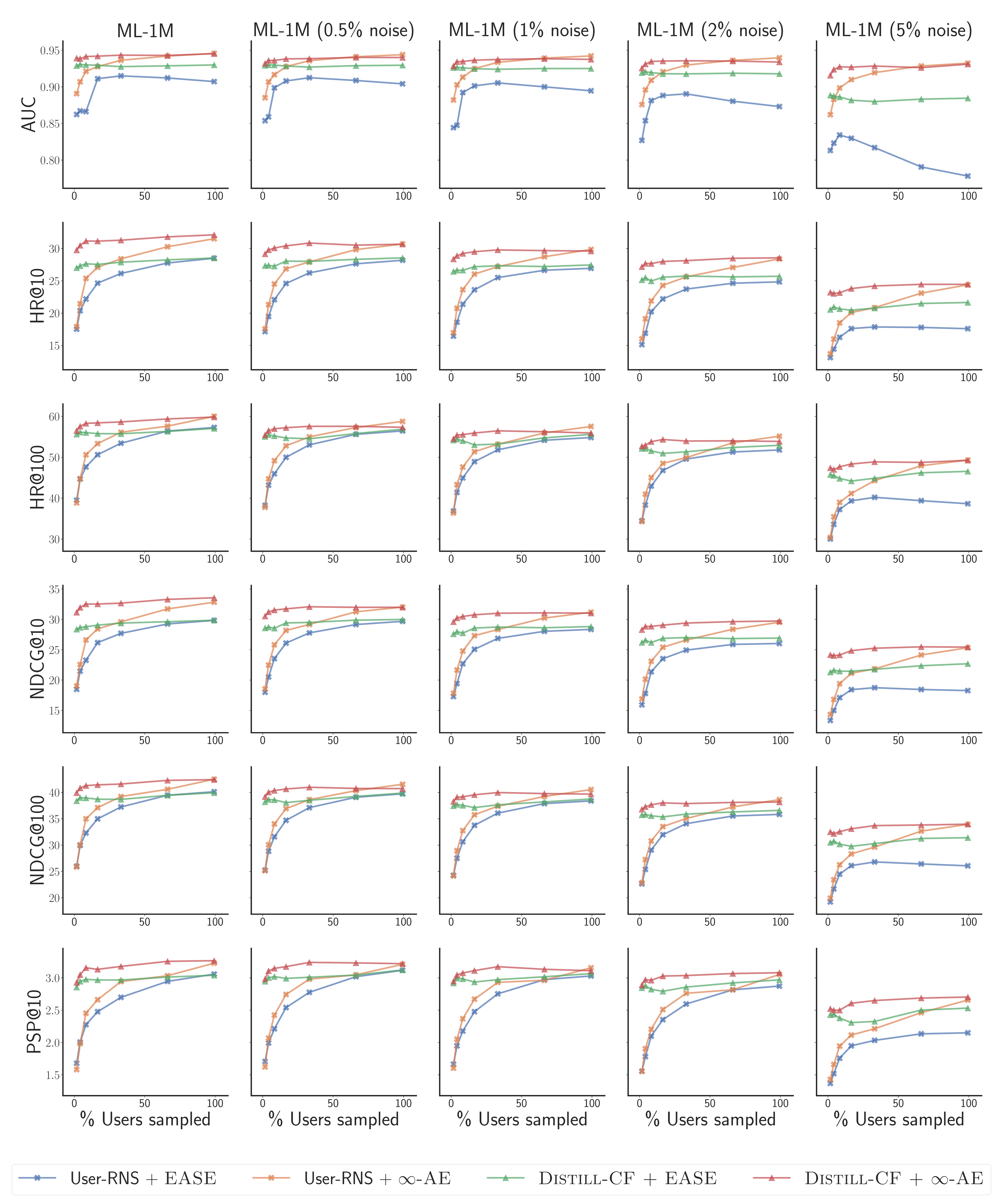}
    \caption{Performance of \model on data sampled by \sampler and User-RNS when there's noise in the data. Results for EASE have been added for reference. Each column represents a specific level of noise in the original data, and each row represents an evaluation metric. All results are on the MovieLens-1M dataset.}
    \label{fig:denoise_sample_complexity_all_metrics}
\end{figure}

\paragraph{Additional plots on the robustness of \sampler \& \model to noise.} In addition to \cref{fig:denoise_sampler} and \cref{fig:denoise_sample_complexity} in the main paper, we plot results for the EASE model trained on data sampled by different sampling strategies, when there's varying levels of noise in the original data. We plot this for the MovieLens-1M dataset and all metrics in \cref{fig:inf_ae_sampling_all_metrics}. We notice similar trends for all metrics across datasets. We also plot the sample complexity results for EASE and \model over the MovieLens-1M dataset and all metrics in \cref{fig:denoise_sample_complexity_all_metrics}. We observe similar trends across metrics.

\begin{figure}[ht!] 
    \includegraphics[width=\linewidth]{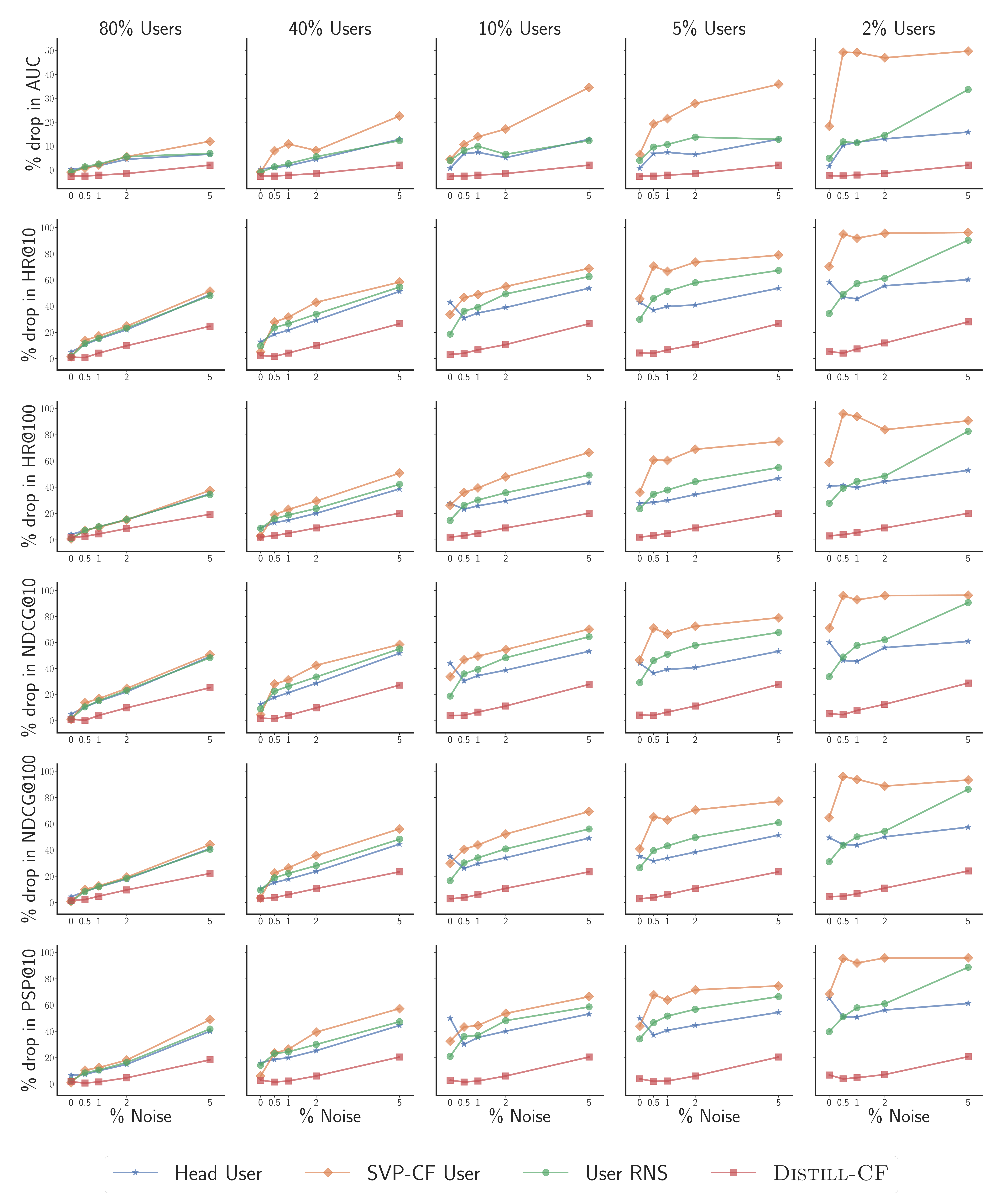}
    \caption{Performance of the EASE model trained on data sampled by different sampling strategies when there's varying levels of noise in the data. Each column represents a user sampling budget, and each row represents the \% drop w.r.t a single evaluation metric. All results are on the MovieLens-1M dataset.}
    \label{fig:denoise_sampler_all_metrics}
\end{figure}

\end{document}

%% file: general_definitions.tex
\makeatletter \renewcommand\paragraph{\@startsection{paragraph}{4}{\z@} {2mm \@plus1ex \@minus.2ex} {-0.7em} {\normalfont\normalsize\bfseries}} \makeatother

\newcommand{\listheader}[1]{\item \emph{#1} }

\newcommand\overstar[1]{\ThisStyle{\ensurestackMath{%
  \setbox0=\hbox{$\SavedStyle#1$}%
  \stackengine{0pt}{\copy0}{\kern.2\ht0\smash{\SavedStyle*}}{O}{c}{F}{T}{S}}}}

\newtheorem{theorem}{Theorem}[section]
\newtheorem{lemma}[theorem]{Lemma}
\newtheorem{proposition}[theorem]{Proposition}
\newtheorem{definition}[theorem]{Definition}

\newcommand{\STAB}[1]{\begin{tabular}{@{}c@{}}#1\end{tabular}}

\newcommand{\bs}[1]{\ensuremath{\bm{\mathit{#1}}}}

\newcommand{\ie}[0]{\emph{i.e.}\xspace}
\newcommand{\etc}[0]{\emph{etc.}\xspace}
\newcommand{\eg}[0]{\emph{e.g.}\xspace}
\newcommand{\vs}[0]{\emph{vs.}\xspace}
\newcommand{\wrt}[0]{\emph{w.r.t.}\xspace}

%% file: recsys_definitions.tex
\newcommand{\dataset}[0]{$\mathcal{D}$\xspace}
\newcommand{\sampled}[0]{$\mathcal{D}'$\xspace}
\newcommand{\argmin}[1]{\underset{#1}{\operatorname{arg}\,\operatorname{min}}\;}
\newcommand{\expectation}[2]{\underset{#1}{\mathbb{E}} \left[#2\right]}

\newcommand{\sampler}{\textsc{Distill-CF}\xspace}
\newcommand{\model}{\textsc{$\infty$-AE}\xspace}

\newcommand{\EE}{\operatornamewithlimits{\mathbb{E}}} 

%% file: tables/model_results.tex
\newcommand{\bb}[1]{\textbf{#1}}
\newcommand{\uu}[1]{\underline{#1}}
\begin{table*}
    \begin{scriptsize} 
    \caption{Comparison of \model with different methods on various datasets. All metrics are better when higher. Brief set of data statistics can be found in \cref{appendix:hyper_params}, \cref{data_stats}. \bb{Bold} values represent the best in a given row, and \uu{underlined} represent the second-best. Results for \model on the Netflix dataset (marked with a *) consist of random user-sampling with a max budget of $25$K $\equiv 5.4$\% users, and results for \sampler + \model have a user-budget of $500$ for all datasets.}
    \label{tab:results}
    \begin{center}
        \begin{tabular}{c | c | c c c c c c c | c c}
            \toprule
            \multirow{2}{*}{Dataset} & \multirow{2}{*}{Metric} & \multirow{2}{*}{PopRec} & \multirow{2}{*}{\STAB{Bias\\only}} & \multirow{2}{*}{MF} & \multirow{2}{*}{NeuMF} & \multirow{2}{*}{\STAB{Light\\GCN}} & \multirow{2}{*}{EASE} & \multirow{2}{*}{MVAE} & \multirow{2}{*}{\model} & \multirow{2}{*}{\STAB{\sampler\\+ \model}} \\
            
            & & & & & & & & & \\
            
            \midrule
            
            \multirow{6}{*}{\textbf{\STAB{Amazon\\Magazine}}} 
            & AUC       & 0.8436    & 0.8445    & 0.8475    & 0.8525    & 0.8141    & 0.6673        & 0.8507    & \uu{0.8539}   & \bb{0.8584} \\
            & HR@10     & 14.35     & 14.36     & 18.36     & 18.35     & \uu{27.12}& 26.31         & 17.94     & 27.09         & \bb{28.27} \\
            & HR@100    & 59.5      & 59.35     & 58.94     & 59.3      & 58.00     & 48.36         & 57.3      & \uu{60.86}    & \bb{61.78} \\
            & NDCG@10   & 8.42      & 8.33      & 13.1      & 13.6      & 22.57     & 22.84         & 12.18     & \uu{23.06}    & \bb{23.81} \\
            & NDCG@100  & 19.38     & 19.31     & 21.76     & 21.13     & 29.92     & 28.27         & 19.46     & \uu{30.75}    & \bb{31.75} \\
            & PSP@10    & 6.85      & 6.73      & 9.24      & 9.00      & 13.2      & 12.96         & 8.81      & \uu{13.22}    & \bb{13.76} \\
            
            \midrule
            
            \multirow{6}{*}{\textbf{ML-1M}} 
            & AUC       & 0.8332    & 0.8330    & 0.9065    & 0.9045    & 0.9289    & 0.9069    & 0.8832    & \bb{0.9457}   & \uu{0.9415} \\
            & HR@10     & 13.07     & 12.93     & 24.63     & 23.25     & 27.43     & 28.54     & 21.7      & \bb{31.51}   & \uu{31.16} \\
            & HR@100    & 30.38     & 29.63     & 53.26     & 51.42     & 55.61     & 57.28     & 52.29     & \bb{60.05}   & \uu{58.28} \\
            & NDCG@10   & 13.84     & 13.74     & 25.65     & 24.44     & 28.85     & 29.88     & 22.14     & \bb{32.82}   & \uu{32.52} \\
            & NDCG@100  & 19.49     & 19.13     & 35.62     & 33.93     & 38.29     & 40.16     & 33.82     & \bb{42.53}   & \uu{41.29} \\
            & PSP@10    & 1.10      & 1.07      & 2.41      & 2.26      & 2.72      & 3.06      & 2.42      & \bb{3.22}   & \uu{3.15} \\
            
            \midrule
            
            \multirow{6}{*}{\textbf{Douban}} 
            & AUC       & 0.8945    & 0.8932    & 0.9288    & 0.9258    & 0.9391    & 0.8570    & 0.9129    & \bb{0.9523}   & \uu{0.9510} \\
            & HR@10     & 11.06     & 10.71     & 12.69     & 12.79     & 15.98     & 17.93     & 15.36     & \bb{23.56}   & \uu{22.98} \\
            & HR@100    & 17.07     & 16.63     & 20.29     & 19.69     & 22.38     & 25.41     & 22.82     & \bb{28.37}   & \uu{27.20} \\
            & NDCG@10   & 11.63     & 11.24     & 13.21     & 13.33     & 16.68     & 19.48     & 16.17     & \bb{24.94}   & \uu{24.20} \\
            & NDCG@100  & 12.63     & 12.27     & 14.96     & 14.39     & 17.20     & 19.55     & 17.32     & \bb{23.26}   & \uu{22.21} \\
            & PSP@10    & 0.52      & 0.50      & 0.63      & 0.63      & 0.86      & 1.06      & 0.87      & \bb{1.28}   & \uu{1.24} \\
            
            \midrule
            
            
            
            \multirow{6}{*}{\textbf{Netflix}} 
            & AUC       & 0.9249    & 0.9234    & 0.9234    & 0.9244    & \multirow{6}{*}{\STAB{Timed\\Out}}   & 0.9418   & 0.9495    & \uu{0.9663}*   & \bb{0.9728} \\
            & HR@10     & 12.14     & 11.49     & 11.69     & 11.06     &    & 26.03        & 20.6     & \bb{29.69}*    & \uu{29.57} \\
            & HR@100    & 28.47     & 27.66     & 27.72     & 26.76     &    & \uu{50.35}   & 44.53    & \bb{50.88}*    & 49.24 \\
            & NDCG@10   & 12.34     & 11.72     & 12.04     & 11.48     &    & 26.83        & 20.85    & \bb{30.59}*    & \uu{30.54} \\
            & NDCG@100  & 17.79     & 16.95     & 17.17     & 16.40     &    & 35.09        & 29.22    & \bb{36.59}*    & \uu{35.58} \\
            & PSP@10    & 1.45      & 1.28      & 1.31      & 1.21      &    & 3.59         & 2.77     & \bb{3.75}*     & \uu{3.62} \\
            
            \bottomrule
        \end{tabular}
    \end{center}
    \end{scriptsize}
\end{table*}

%% file: algorithms/inf_ae_train.tex
\vspace{-0.5cm}
\begin{algorithm}[H]
\centering
\caption{\model model training} \label{alg:inf_ae_trian}
    \begin{flushleft}
        \textbf{Input:} $\text{User set}~~ \mathcal{U} ~;~ \text{dataset}~ \mathbf{X} \in \mathbb{R}^{|\mathcal{U}| \times |\mathcal{I}|} ;$ NTK \\
        $\mathbb{K} : \mathbb{R}^{|\mathcal{I}|} \times \mathbb{R}^{|\mathcal{I}|} \mapsto \mathbb{R} ~;~ \text{regularization const.}$ \\ $\lambda \in \mathbb{R}$ \\
        \textbf{Output:} $\text{Dual parameters}~ \alpha \in \mathbb{R}^{|\mathcal{U}| \times |\mathcal{I}|}$
    \end{flushleft}
    \vspace{0.1cm}
    \begin{algorithmic}[1]
        \Procedure{fit}{$\mathcal{U}, \mathbf{X}, \mathbb{K}$} 
            \State $\mathbf{K} \leftarrow [ 0 ]_{|\mathcal{U}| \times |\mathcal{U}|}$ 
            \Comment{Zero Initialization}
            \State $\mathbf{K}_{u, v} \leftarrow \mathbb{K}(\mathbf{X}_u, \mathbf{X}_v) ~~~ \forall u \in \mathcal{U}, v \in \mathcal{U}$
            \State $\alpha \leftarrow (\mathbf{K} + \lambda I)^{-1} \cdot \mathbf{X}$
            \State \Return $\alpha$
        \EndProcedure
    \end{algorithmic}
    \vspace{0.08cm}
\end{algorithm}

%% file: algorithms/inf_ae_predict.tex
\vspace{-0.5cm}
\begin{algorithm}[H]
\centering
\caption{\model inference} \label{alg:inf_ae_predict}
    \begin{flushleft}
        \textbf{Input:} $\text{User set}~~ \mathcal{U} ~;~ \text{dataset}~ \mathbf{X} \in \mathbb{R}^{|\mathcal{U}| \times |\mathcal{I}|} ;$ NTK \\
        $\mathbb{K} : \mathbb{R}^{|\mathcal{I}|} \times \mathbb{R}^{|\mathcal{I}|} \mapsto \mathbb{R} ;~ \text{dual params.}~ \alpha \in \mathbb{R}^{|\mathcal{U}| \times |\mathcal{I}|}$ \\
        $;~ \text{inference user history}~ \hat{\mathbf{X}}_u \in \mathbb{R}^{|\mathcal{I}|} $ \\
        \textbf{Output:} $\text{Prediction}~ \hat{y} \in \mathbb{R}^{|\mathcal{I}|}$
    \end{flushleft}
    \begin{algorithmic}[1]
        \Procedure{predict}{$\mathcal{U}, \mathbf{X}, \hat{\mathbf{X}}_u, \mathbb{K}, \alpha$} 
            \State $\mathbf{K} \leftarrow [0]_{|\mathcal{U}|}$ 
            \Comment{Zero Initialization}
            \State $\mathbf{K}_{v} \leftarrow \mathbb{K}(\hat{\mathbf{X}}_u, \mathbf{X}_v) ~~~ \forall v \in \mathcal{U}$
            \State $\hat{y} \leftarrow softmax(\mathbf{K} \cdot \alpha)$
            \State \Return $\hat{y}$
        \EndProcedure
    \end{algorithmic}
\end{algorithm}

%% file: algorithms/distillation.tex
\begin{algorithm}[H]
\centering
\caption{Data synthesis using \sampler} \label{alg:distill}
    \begin{flushleft}
        \textbf{Input:} $\text{User set}~~ \mathcal{U} ~;~ \text{dataset}~ \mathbf{X} \in \mathbb{R}^{|\mathcal{U}| \times |\mathcal{I}|} ;$ NTK $\mathbb{K} : \mathbb{R}^{|\mathcal{I}|} \times \mathbb{R}^{|\mathcal{I}|} \mapsto \mathbb{R} ~;~ \text{support user size}~ \mu \in \mathbb{R} ~;~ $ \\
        gumbel softmax temperature $\tau \in \mathbb{R} ~;~ \text{reg. const.}~ \lambda_2 \in \mathbb{R} ~;~ \text{SGD batch-size}~ b, \text{step-size}~ \eta \in \mathbb{R}$ \\
        \textbf{Output:} Synthesized data $\mathbf{X^s} \in \mathbb{R}^{\mu \times |\mathcal{I}|}$
    \end{flushleft}
    \begin{algorithmic}[1]
        \Procedure{sample}{$n, \mathcal{U}, \mathbf{X}$} 
            \State $\mathcal{U}' \sim \mathcal{U}$ \Comment{Randomly sample $n$ users from $\mathcal{U}$}
            \State $\mathbf{X'} \leftarrow \mathbf{X_u} ~~~ \forall u \in \mathcal{U}'$ \Comment{Retrieve corresponding rows from $\mathbf{X}$}
            \State \Return $\mathcal{U}', \mathbf{X'}$
        \EndProcedure
    
        \Procedure{synthesize}{$\mathcal{U}, \mathbf{X}, \mathbb{K}$} 
            \State $\mathcal{U}^s, \mathbf{X^s} \leftarrow \textsc{sample}(\mu, \mathcal{U}, \mathbf{X})$ \Comment{Sample support data}
            \For{$steps = 0 \ldots \xi$}
                \State $\mathbf{\hat{X}^s} \leftarrow \sigma \left[ \sum_{i=1}^{\gamma} gumbel_{\tau}(softmax(\mathbf{X^s})) \right]$
                \State $\alpha^s \leftarrow \textsc{fit}(\mathcal{U}^s, \mathbf{\hat{X}^s}, \mathbb{K})$ 
                \Comment{Fit \model on support data}
                \State $\mathcal{U}^b, \mathbf{X^b} \leftarrow \textsc{sample}(b, \mathcal{U}, \mathbf{X})$
                \State $\mathbf{\Tilde{X}} \leftarrow [0]_{b \times |\mathcal{I}|}$
                \State $\mathbf{\Tilde{X}_u} \leftarrow \textsc{predict}(\mathcal{U}^s, \mathbf{\hat{X}^s}, \mathbf{X^b_u}, \mathbb{K}, \alpha^s) ~~~ \forall u \sim \mathcal{U}^b$ \Comment{Predict for all sampled users}
                \State $L \leftarrow \mathbf{X^b} \cdot log(\mathbf{\Tilde{X}}) + (1 - \mathbf{X^b}) \cdot log(1 - \mathbf{\Tilde{X}}) + \lambda_2 \cdot ||\mathbf{\hat{X}^s}||_1$ \Comment{Re-construction loss}
                \State $\mathbf{X^s} \leftarrow \mathbf{X^s} - \eta \cdot \dfrac{\partial L}{\partial \mathbf{X^s}}$ \Comment{SGD on $\mathbf{X^s}$}
            \EndFor
            \State \Return $\mathbf{X^s}$
        \EndProcedure
    \end{algorithmic}
\end{algorithm}

%% file: tables/data_stats.tex
\begin{table*}
    \begin{small} 
    \caption{Brief set of statistics of the datasets used in this paper.}
    \label{data_stats}
    \begin{center}
        \begin{tabular}{c | c c c c}
            \toprule
            Dataset & \# Users & \# Items & \# Interactions & Sparsity \\
            
            \midrule
            
            \textbf{Amazon Magazine \cite{amz_data}}            & 3k & 1.3k & 12k & 99.7\% \\
            \textbf{ML-1M \cite{movielens}}                      & 6k & 3.7k & 1M & 95.6\% \\
            \textbf{Douban \cite{douban_data}}                     & 2.6k & 34k & 1.2M & 98.7\% \\
            \textbf{Netflix \cite{netflix_data}}                    & 476k & 17k & 100M & 98.9\% \\
            
            \bottomrule
        \end{tabular}
    \end{center}
    \end{small}
\end{table*}

%% file: tables/hyper_params.tex
\begin{table*}
    \begin{small} 
    \caption{List of all the hyper-parameters grid-searched for \model, \sampler, and baselines.}
    \label{tab:hyper_params}
    \begin{center}
        \begin{tabular}{c c | c c c c}
            \toprule
            Hyper-Parameter & Model & Amz. Magazine & ML-1M & Douban & Netflix \\
            
            \midrule
            
            \multirow{4}{*}{Latent size}    & MF        & \multicolumn{4}{c}{\multirow{4}{*}{\{4, 8, 16, 32, 50, 64, 128\}}} \\
                                            & NeuMF     & & & & \\
                                            & LightGCN  & & & & \\
                                            & MVAE      & & & & \\
            
            \midrule
            
            \multirow{5}{*}{\# Layers}      & MF        & \multicolumn{4}{c}{\multirow{4}{*}{\{1, 2, 3\}}} \\
                                            & NeuMF     & & & & \\
                                            & LightGCN  & & & & \\
                                            & MVAE      & & & & \\
                                            & \model    & \multicolumn{4}{c}{\{1\}} \\
            
            \midrule
            
            \multirow{5}{*}{Learning rate}  & MF        & \multicolumn{3}{c}{\multirow{4}{*}{\{0.001, 0.006, 0.01\}}} & \multirow{4}{*}{\{0.006\}} \\
                                            & NeuMF     & & & & \\
                                            & LightGCN  & & & & \\
                                            & MVAE      & & & & \\
                                            & \sampler  & \multicolumn{4}{c}{\{0.04\}} \\
            
            \midrule
            
            \multirow{4}{*}{Dropout}        & MF        & \multicolumn{4}{c}{\multirow{4}{*}{\{0.0, 0.3, 0.5\}}} \\
                                            & NeuMF     & & & & \\
                                            & LightGCN  & & & & \\
                                            & MVAE      & & & & \\
            
            \midrule
            
            \multirow{3}{*}{$\lambda$}      & EASE      & \multicolumn{4}{c}{\{1, 10, 100, 1K, 10K\}} \\
                                            & \model    & \multicolumn{4}{c}{\{0.0, 1.0, 5.0, 20.0 50.0, 100.0\}} \\
                                            & \sampler  & \multicolumn{4}{c}{\{1e-5, 1e-3, 0.1, 1.0, 5.0, 50.0\}} \\
            
            \midrule
            
            $\lambda_2$                     & \sampler  & \multicolumn{4}{c}{$\dfrac{\text{\{1e-3, 10.0\}}}{\text{avg. \# of interactions per user}}$} \\
            
            \midrule
            
            $\tau$                          & \sampler  & \multicolumn{4}{c}{\{0.3, 0.5, 0.7, 5.0\}} \\
            
            \midrule
            
            $\gamma$                        & \sampler  & \{50, 100, 200\} & \{200, 500, 700\} & \{500, 1K, 2K\} & \{500, 700\} \\
            
            \bottomrule
        \end{tabular}
    \end{center}
    \end{small}
\end{table*}